\documentclass[journal]{IEEEtran}

\usepackage{url}
\usepackage[most]{tcolorbox}
\usepackage{algorithmic}
\usepackage{algorithm}
\usepackage{array}
\usepackage{amssymb}
\usepackage{booktabs}  
\usepackage{stfloats}
\usepackage{pifont}
\usepackage[T2A]{fontenc}
\usepackage[utf8]{inputenc}
\usepackage[russian,english]{babel}

\usepackage{xcolor} 
\usepackage{textcomp}
\usepackage{verbatim}
\usepackage{graphicx}
\usepackage{cite}
\usepackage{hyperref}
\usepackage{subcaption}
\usepackage{multirow}
\usepackage{booktabs}
\usepackage{makecell}
\usepackage{siunitx}
\usepackage{graphicx}
\usepackage{tikz}
\usepackage{array} 
\usepackage{xtab}
\usepackage{balance}
\usepackage{comment}
\usepackage{tikz}
\usetikzlibrary{shapes.geometric, arrows.meta, positioning}

\newcolumntype{P}[1]{>{\centering\arraybackslash}p{#1}}

\begin{document}

\title{A Comprehensive Survey on Synthetic Network Traffic Generation}

\author{Nirhoshan Sivaroopan,~
       Kaushitha Silva,~
        Chamara Madarasingha,~
        Thilini Dahanayaka,\\
        Guillaume Jourjon,~
         Anura Jayasumana,~
        and~Kanchana Thilakarathna

\thanks{Nirhoshan Sivaroopan (e-mail: \href{mailto:nirhosh98@gmail.com}{snir5742@uni.sydney.edu.au}), Thilini Dahanayaka (e-mail: \href{mailto:thilini.dahanayaka@sydney.edu.au}{thilini.dahanayaka@sydney.edu.au}), Kanchana Thilakarathna (e-mail: \href{mailto:kanchana.thilakarathna@sydney.edu.au}{kanchana.thilakarathna@sydney.edu.au})  are with the University of Sydney, Australia.}
\thanks{Kaushitha Silva (e-mail: \href{mailto:e19372@eng.pdn.ac.lk}{e19372@eng.pdn.ac.lk}) is with University of Peradeniya, Sri Lanka .}
\thanks{Chamara Madarasingha (e-mail: \href{mailto:c.kattadige@curtin.edu.au}{c.kattadige@curtin.edu.au})is with Curtin University, Australia .}
\thanks{Guillaume Jourjon (e-mail: \href{mailto:guillaume.jourjon@data61.csiro.au}{guillaume.jourjon@data61.csiro.au}) is with CSIRO, Australia .}
\thanks{Anura Jayasumana (e-mail: \href{Anura.Jayasumana@colostate.edu}{Anura.Jayasumana@colostate.edu}) is with Colorado State University, USA.}

}

\maketitle

\begin{abstract}

Synthetic network traffic generation has emerged as a promising alternative for various data-driven applications in the networking domain. It enables the creation of synthetic data that preserves real-world characteristics while addressing key challenges such as data scarcity, privacy concerns, and purity constraints associated with real data. In this survey, we provide a comprehensive review of synthetic network traffic generation approaches, covering essential aspects such as data types and generation models. With the rapid advancements in Artificial Intelligence (AI) and Machine Learning (ML), we focus particularly on deep learning (DL)-based techniques while also providing a detailed discussion of statistical methods and their extensions, including commercially available tools. We present a comprehensive comparision of generation approaches and provide an AI tool to apply this comparision for any network traffic generation papers. Furthermore, we highlight open challenges in this domain and discuss potential future directions for further research and development. This survey serves as a foundational resource for researchers and practitioners, offering a structured analysis of existing methods, challenges, and opportunities in synthetic network traffic generation.

\end{abstract}


\IEEEpeerreviewmaketitle

\vspace{-2mm}
\section{Introduction}
With the unprecedented growth of network traffic and the variety of network-based applications in recent years, traffic models and traffic traces have become essential components for both network providers and users for numerous tasks. 
For example, accurate and representative traffic plays a significant role in understanding network behavior~\cite{xu2008internet, kim2016enabling}, maintaining Quality of Service (QoS)~\cite{liu2008novel, badawy2020qos, mazhar2023quality}, optimizing systems~\cite{ricciato2006traffic, wang2010traffic}, and benchmarking critical network systems~\cite{parsonson2022traffic}. Furthermore, network traffic analysis is crucial for securing network systems while protecting user privacy~\cite{kattadige2021seta++, wang2020fingerprinting, satam2015anomaly}, identifying bottlenecks in real-time~\cite{alzibdeh2023bandwidth, calyam2005tbi}, and forecasting future demands~\cite{9349573, cortez2006internet, papagiannaki2005long} to name a few. However, the need to  depend on learning-based strategies to tackle such issues under complex scenarios has renewed interest in synthetic traffic generation. 

\vspace{-2mm}
\subsection{Need for accurate network traffic data and challenges}
While measured traffic traces and those generated using statistical models served as a main pillar of network traffic analysis in the past, the emerging design, provisioning, and management tools are based on data-driven approaches such as Machine Learning, Artificial Intelligence  and simulated environments such as Digital Twin (DT). A major challenge for developing and training such models is the need for large and complex datasets that represent network traffic including the interaction of myriads of technologies, protocols, topologies,  and applications. For example, Deep Neural Network (DNN)–based tools typically require large volumes of network traffic data to achieve optimal performance~\cite{sivaroopan2024netdiffus, kattadige2021seta++}. Similarly, DT technology—an enabler for simulating 5G/6G networking in industrial Internet of Things (IoT), testing network disruptions and developing Vehicle-to-Everything (V2X) systems~\cite{mihai2022digital, li2024digital} etc., relies on accurate network traffic data even before real-world implementations.

Even with dedicated systems and tools, collecting high-quality network traffic from real-world settings remains challenging due to multiple reasons.
First, the sheer volume of data transmitted through networks makes it difficult to extract relevant traffic while filtering out noisy background data.
In particular, these networks and the related traffic are highly diverse in their characteristics (e.g., protocol, burstiness, packet sizes etc.). For example, wired backbone networks often display high-throughput and relatively stable flows, whereas wireless and cellular networks are influenced by factors such as signal variability, interference, and user mobility, leading to more bursty and dynamic traffic patterns.

Second, deploying data collection tools in complex network environments often requires extended periods, leading to delays in completing  network traffic analysis tasks~\cite{clegg2009challenges}. 
Third, controlling the traffic sources to operate a network under the necessary conditions for trace gathering or for creating anomalies may interfere with normal operation or even be impossible to achieve. Additionally, stringent data regulations and privacy policies  restrict operators and researchers from freely collecting, sharing, or analyzing traffic data~\cite{porras2006large, guerra2022datasets}.

\vspace{-2mm}
\subsection{Significance of synthetic network traffic generation}

To address these challenges and support diverse network applications, researchers have increasingly turned to synthetic network traffic generation, a trend further motivated by the success of synthetic data in other domains. For instance, synthetic data has been widely adopted in medical image–based diagnosis~\cite{zhou2024privacy, chen2021synthetic}, object detection for autonomous vehicles~\cite{kiefer2022leveraging}, and fault detection in industrial systems~\cite{wu2019faultseg3d, boikov2021synthetic}.

The primary aim of synthetic network traffic generation frameworks and models is to create artificial datasets that preserve the key characteristics of real-world traffic. These frameworks are typically based on mechanisms such as statistical approaches, deep learning–based methods, or commercially developed traffic simulation tools, each capable of generating diverse types of network traffic data. By leveraging synthetic data, either independently or alongside limited real-world traces, many network traffic analysis tasks have demonstrated enhanced performance~\cite{sivaroopan2023synig, sivaroopan2024netdiffus, kattadige2021videotrain}, highlighting the value of synthetically generated data. 
For example, it has been already proven that many downstream network traffic analytics tasks including ML-based network traffic classification~\cite{sivaroopan2023synig, madarasingha2022videotrain++}, traffic modeling~\cite{arfeen2019role}, network security related activities~\cite{shi2019generative,qu2024towards}, and network simulation development~\cite{yang2024research,kim2024network} etc. have been benefited through synthetic traffic data. To support these requirements, synthetic network traffic generation acquire positive attributes in various aspects as below.  
\vspace{-5mm}

\begin{itemize}
    \item Data fidelity: Modern generative AI models (e.g., DNN-based) can capture subtle variations in network traffic and generate synthetic data that closely mirrors the properties of real-world traffic.  
    \item Data diversity: These models can produce network traffic for a wide range of network data types, e.g., from local area networks (LAN) to wired backbone networks.  
    \item Safety and security: Enables controlled emulation of rare failures, attacks, and anomaly conditions for training and evaluating defense mechanisms without disrupting production networks.
    \item Benchmarking and reproducibility: Provides standardized, shareable datasets to compare algorithms fairly across research groups and vendors.
    \item Pre-deployment validation: Supports stress testing of new architectures (e.g., 6G, IoT, V2X) and closed-loop Artificial Intelligence for IT Operations (AIOps) pipelines before real traffic is available.
    \item Privacy and compliance: Facilitates experimentation while avoiding exposure of sensitive user data and easing compliance with data protection regulations.
    \item Cost and coverage: Reduces data collection costs and enables exploration of corner cases that are rare or infeasible to capture in the wild.
    \item Time efficiency: Compared to deploying hardware or software data collection tools, which require prolonged time, these models can be trained with  limited amount of already available data relatively taking less time. 
\end{itemize}

Reviewing the recent research literature, we observe that a plethora of work has been done in this domain, which we identify under two primary categories: \textit{i})~\textit{Statistical models}, which analyze the statistical characteristics of traffic features and generate new data that conform to these properties. \textit{ii})~\textit{DL-based models} (Variational Autoencoders (VAEs), Generative Adversarial Networks (GANs), Diffusion Models (DMs), and transformer-based techniques) supported by DNNs. These models have shown remarkable performance in learning subtle and complex traffic patterns to generate high-fidelity data.
Extending the models in these two approaches, \textit{network traffic simulators} exist that generate network packet streams while preserving protocol-specific attributes. Furthermore, \textit{commercial traffic generators}, with configurable parameters are applied in both research and industrial settings. 

\begin{figure}[h!]
\centering
\includegraphics[width=1\columnwidth]{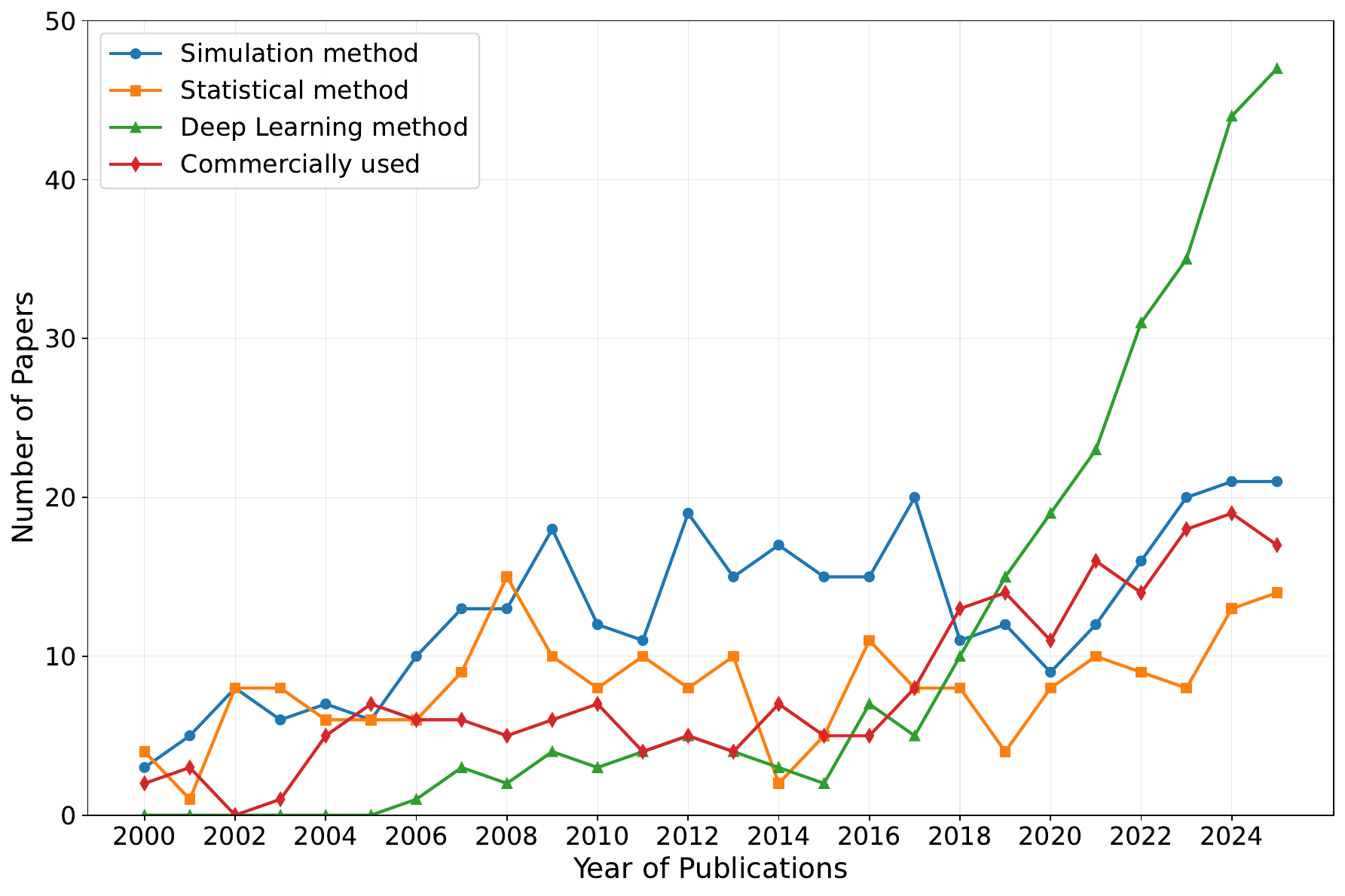}
\caption{Research papers in network traffic generation. The numbers were extracted by querying the Scopus database.}
\label{fig:paper_trend}
\end{figure}

To highlight the growing interest and, hence, the significance in the field of network traffic generation, Fig.~\ref{fig:paper_trend} presents the number of publications in this domain from 2000 to 2024 under the aforementioned categories.
The articles were retrieved through a Scopus search using relevant keywords in the titles, abstracts, and keyword sections. 
we observe a sustained increase in publications across all categories over the last two decades, with the steepest growth in DL-based methods after 2018. Simulation- and statistical-based approaches remain active but comparatively flatter, indicating methodological maturity and the inadequate scalability and configurability to meet the emerging needs, while the steady rise of commercially used tools since the mid-2010s signals growing industrial adoption. The overall peak in 2023--2024 underscores that the importance of traffic generation is increasingly recognized by both researchers and practitioners.

\subsection{Related Surveys}\label{sec:related}

Recently, a growing body of work has explored synthetic data generation across diverse domains, including several surveys that examine common challenges in data-driven generation.
Surveys by Goyal \textit{et al.}~\cite{goyal2024systematic}, Bengesi \textit{et al.}~\cite{bengesi2024advancements}, and Figueira \textit{et al.}~\cite{figueira2022survey} examine the topic from a general ML perspective, addressing cross-cutting challenges such as data scarcity, privacy, and algorithmic bias rather than focusing on specific applications. The surveyed models span traditional ML techniques to advanced DL approaches such as GANs, VAEs, diffusion models, and Generative Pre-trained Transformers (GPTs). Figueira \textit{et al.}~\cite{figueira2022survey} provide a detailed overview of GAN-based data generation for tabular datasets, highlighting key advancements, evaluation strategies, and training challenges.

\begin{table*}[t]
\scriptsize
\centering
\caption{Comparison of existing survey papers on synthetic network traffic generation}
\label{tab:survey_comparison}
\begin{tabular}{|p{0.6cm}|p{3.8cm}|p{5.5cm}|p{6.2cm}|}
\hline
\textbf{Survey} & \textbf{Methodologies or Models Covered} & \textbf{Scope of Survey} & \textbf{Limitations Identified} \\
\hline
~\cite{brophy2023generative} & GANs & Time-series modeling of network traffic & Limited to GANs and time-series traffic patterns \\
\hline
~\cite{bourou2021review} & GANs & Intrusion Detection System (IDS) dataset generation using tabular data & Does not generalize beyond tabular attack simulation \\
\hline
~\cite{anande2022generative} & GANs & Evolution and application of GANs for network flows & No coverage of other generative methods \\
\hline
~\cite{agrawal2024review} & GANs & GAN-based attack simulation for cybersecurity testing & GAN-centric, no discussion on evaluation or metrics \\
\hline
~\cite{halvorsen2024applying} & GANs, Autoencoders, Hybrid Models & Generative models for IDS & Broader model types, but focus still limited to IDS context \\
\hline
~\cite{peppes2023comparison} & GAN Variants & Comparative study of GANs for cyberattack data & Focused only on malicious attack traffic and GANs \\
\hline
~\cite{adeleke2022network} & Simulation tools, GANs & Functional categorization of traffic generators & No deep technical breakdown of DL-based generators \\
\hline

\end{tabular}
\end{table*}

Collectively, these studies establish a strong methodological foundation for synthetic data generation. Building on this foundation, numerous works have focused on specific applications such as image generation~\cite{mumuni2024survey} and time-series generation~\cite{iglesias2023data} etc. In the context of network traffic generation, the related surveys (see Table~\ref{tab:survey_comparison}) have primarily examined methodologies that apply generative AI models to capture complex temporal and behavioral patterns in network environments. We note that the majority of existing surveys on network traffic generation are limited to GAN-based approaches, although they span diverse applications such as cyberattack simulation and generic time-series traffic modeling etc.


\subsection{Contributions of this survey}

Motivated by these findings and the growing interest in the field, this survey paper provides a detailed review of network traffic generation. To our knowledge, it is the first survey to systematically explore and compare a wide range of synthetic network traffic generation techniques. 

In contrast to tprevious surveys on network traffic generation, which have primarily focused on specific model types, particularly the GANs, this review broadens the scope by analyzing a diverse range of generative models, including statistical methods, DL approaches, and simulation-based platforms. In particular, we cover recent advancements with state-of-the-art (SOTA) generative models, such as DMs and Transformers, which have not been addressed in previous surveys. 
We also review different network traffic type data generation to provide broader understanding about the nature of data handling with generative models. In addition to existing literature, we discuss open challenges (e.g., privacy-utility tradeoff, complexity, scalability etc.) in network traffic generation with plausible solutions as future research directions.

In this review, we adopt a network- and application-agnostic approach, where the collected network traffic can originate from diverse environments (e.g., LANs, enterprise systems, IoT networks) and support various applications (e.g., cybersecurity, addressing class imbalance). Since most of these networks follow the standard layered architecture, the corresponding data types and protocols (e.g., payload, packet- and flow-level data, physical-layer signals) are similar, and thus the traffic generation process remains consistent.
To summarize, our contributions are as follows:
\begin{itemize}

    \item \textbf{Comprehensive review across different methods:} We provide a detailed analysis of various network traffic generation approaches, from statistical methods and DL-based models to extensions into network traffic simulators and commercially available traffic generators. We provide a broad picture of evolvement of the models in terms of their architectures used in data generation.
    
    \item \textbf{Comparative analysis of major models:} We provide a comparison between major DL-based models using a range of metrics, further guiding users for appropriate model selection for their generative tasks.
    
    \item \textbf{Analysis of different data types used:} 
    We provide a detailed taxonomy of different data types of interest for network traffic generation. This analysis is particularly based on what information resides in data samples and how the real data are collected at different network layers.

    \item \textbf{Identification of open challenges:} We also identify several open challenges, particularly regarding aspects such as privacy concerns, computational efficiency and complex protocol standards, etc. This discussion provides insights for further growth in research in this field.

    \item \textbf{Exploration of future research directions:} Following the open challenges, we provide plausible solutions to overcome those challenges as a guidance to the research community to advance further research.  
    

\end{itemize}

\begin{figure}[h!]
\centering
\includegraphics[width=0.9\columnwidth]{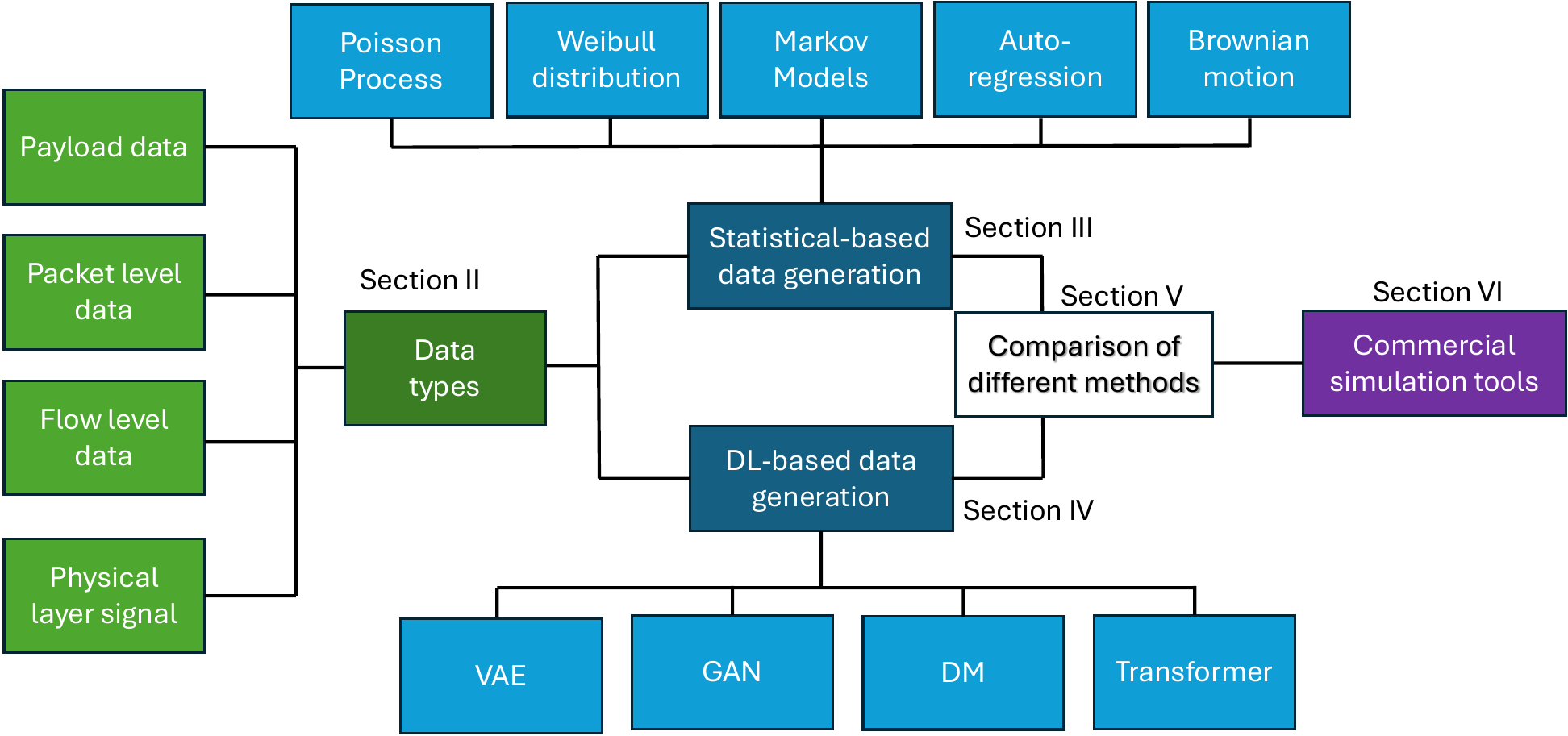}
\caption{Overview of the sections}
\label{fig:sec_overview}
\end{figure}

The remainder of the paper is organized as follows: as shown in Fig.~\ref{fig:sec_overview}, we organize existing techniques into a systematic taxonomy with a structured flow between them. In Section~\ref{sec:data}, we explain the different data types primarily used for network traffic generation, which influence the design and development of the underlying mechanisms (e.g., data pre- and post-processing) in the overall generation pipelines.

Building on these data sources, we first introduce statistical approaches for network traffic generation in Section~\ref{sec:stat}. After highlighting their limitations, we present the DL-based approaches in Section~\ref{sec:dl}, which in general are more advanced tools for network traffic generation that address limitations of statistical methods. We subsequently analyze the distinctions between statistical and deep learning–based approaches and outline a constraint-driven strategy for model selection in Section~\ref{sec:statsvsdl}. Leveraging the knowledge from previous sections, we then describe simulation-based methods in Section~\ref{sec:commercial}. Commonly employed in commercial settings and incorporate both statistical and DL-based techniques as internal mechanisms. Finally, in Section~\ref{sec:challenges_and_future_directions}, we discuss open challenges in network traffic synthesis and provide insights for future research directions, and then conclude the paper in Section~\ref{sec:conclusion}.

\section{Data types in network traffic generation}\label{sec:data}

\begin{figure*}[h!]
\centering
\includegraphics[width=0.9\textwidth]{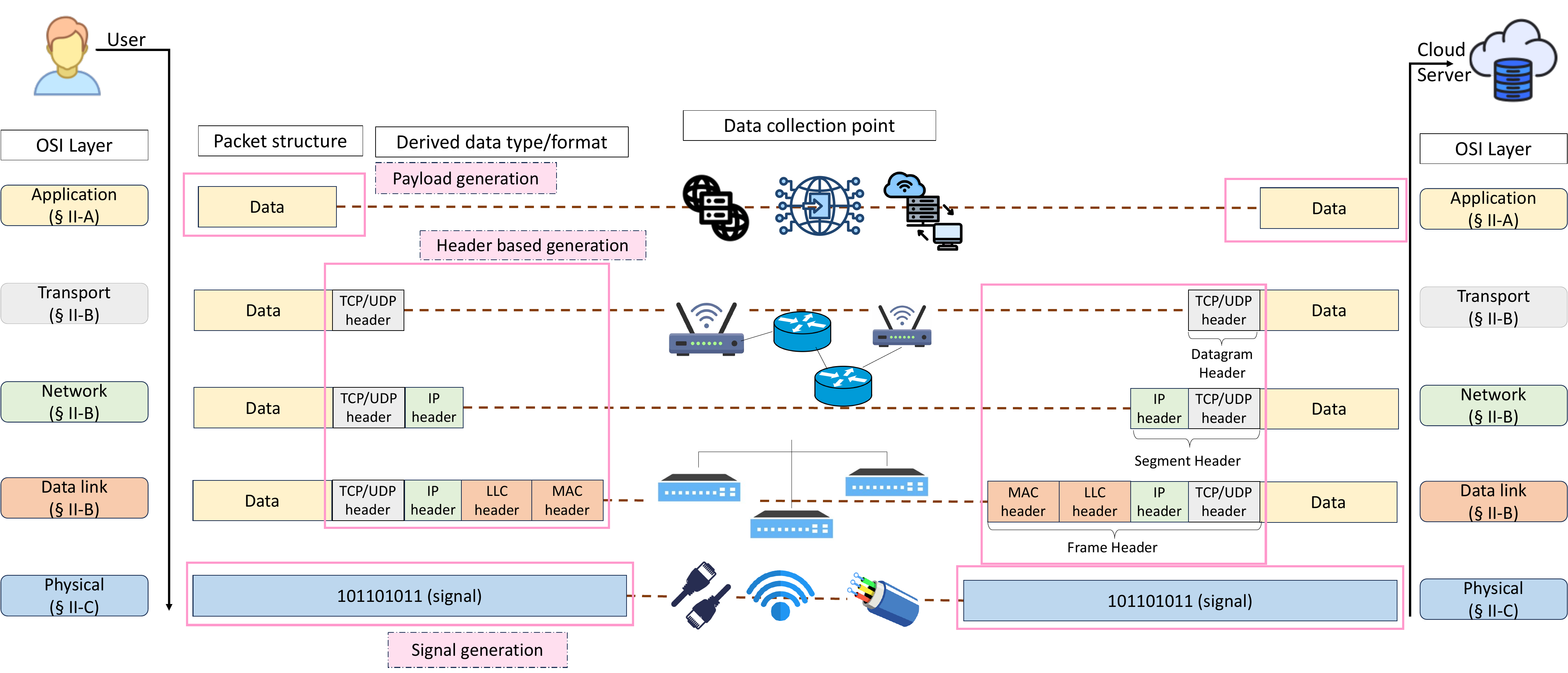}
\vspace{-2mm}
\caption{Network traffic data representations and data collection points organized according to the OSI model. Each layer exposes distinct observable data that can be used for traffic generation at packet or flow granularity.}
\label{fig:data-draft}
\end{figure*}

\begin{figure}[h!]
\centering
\includegraphics[width=1\columnwidth]{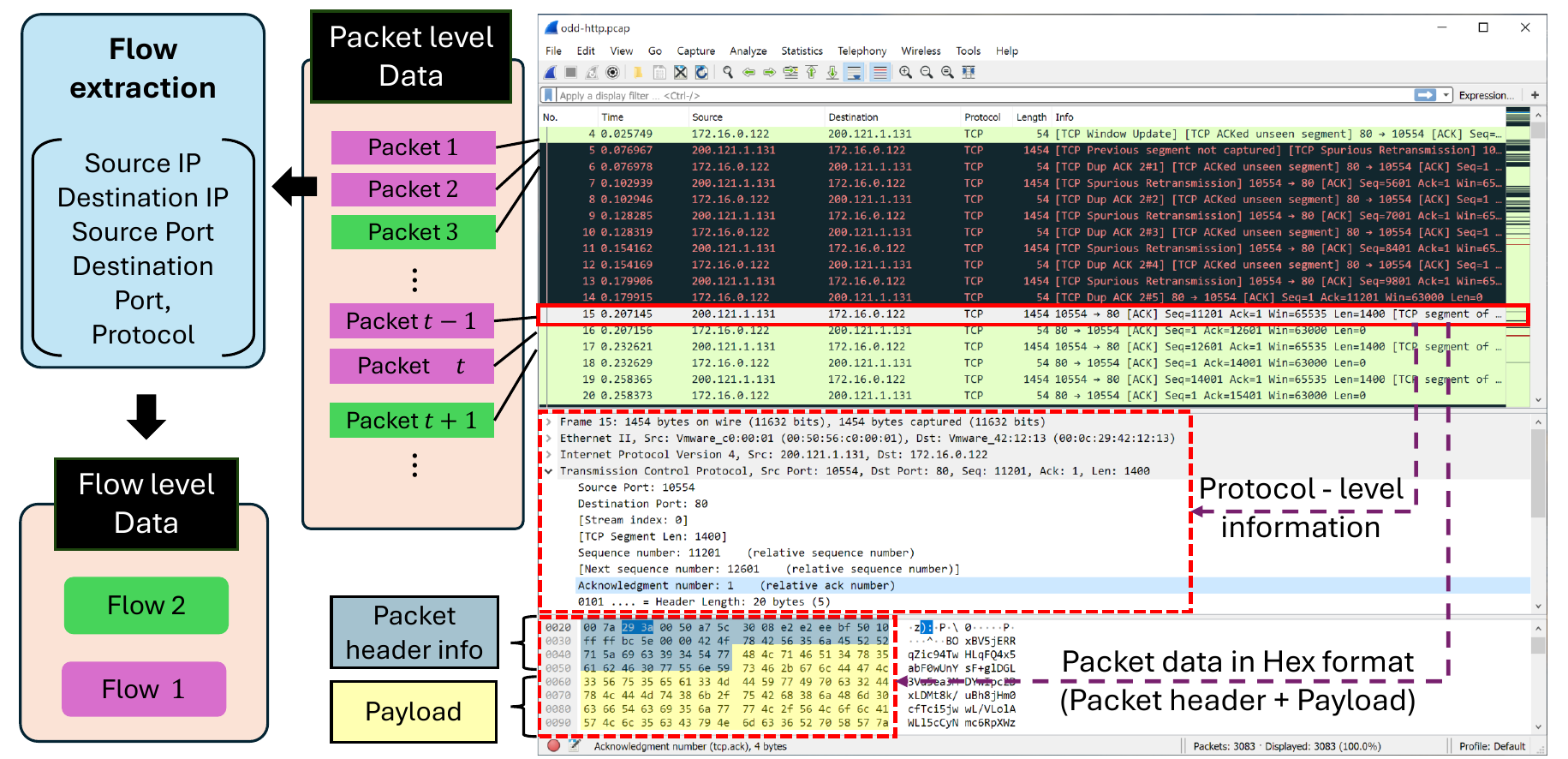}
\caption{Extraction of packet- and flow-level representations from a \texttt{.pcap} file.}
\label{fig:pcap-2data}
\end{figure}

In this section, we characterize the network traffic data used for synthetic traffic generation. Traffic collected from different network vantage points provides different levels of data visibility. For instance, application-layer proxies observe application-level information, often in encrypted form, whereas network routers expose IP addresses, port numbers, and other header fields. At the physical medium, monitoring captures raw physical signals, such as binary sequences or electrical or wireless signals. Consequently, monitoring locations across the network stack expose different subsets of traffic information, leading to substantial variation in the data available for synthetic traffic generation.

To systematically organize these observations, we map each data representation to the Open Systems Interconnection (OSI) model, which clearly characterizes observable network information according to monitoring vantage points. Existing traffic generation models operate on data extracted at different OSI layers and at varying levels of aggregation. Accordingly, we structure the discussion layer by layer, from the Application layer (L7) to the Physical layer (L1). We further distinguish between packet-level and flow-level representations as alternative aggregation views over the same underlying observations. Fig.~\ref{fig:data-draft} summarizes the relationship between network vantage points, OSI layers, observable data types, and common data collection mechanisms.


\subsection{Application layer (L7): Payload and application semantics}

At the Application layer, network traffic is represented by the payload, i.e., the sequence of bytes generated by applications, encrypted and encapsulated within packets. In packet capture tools such as Wireshark~\cite{wireshark} and Tcpdump~\cite{tcpdump}, these bytes are typically visualized in hexadecimal format (Fig.~\ref{fig:pcap-2data}). From a data visibility perspective, full payload access depends strongly on the monitoring vantage point. When traffic is collected at end hosts, proxies, or application gateways, payload contents may be directly observable. In contrast, when monitoring is performed at intermediate network points such as routers, switches, or backbone links, payloads are increasingly encrypted, limiting visibility to encrypted byte streams or metadata only. As this survey focuses on traffic generation from the perspective of these intermediate nodes, the analysis of unencrypted application data is considered out of scope.

Across the literature, application-layer payload data has received comparatively limited attention in traffic generation studies. This is primarily due to two factors. First, modern network traffic is predominantly encrypted (e.g., HTTPS, QUIC), which prevents direct observation of application content at most network vantage points. As a result, realistic payload data can typically only be obtained through end-device instrumentation, trusted middleboxes, or controlled testbed environments. Second, even when available, payload data exhibits high variability and complex semantic structure, making it challenging to model and often unnecessary for many downstream tasks, where header-level features such as packet size, timing, and protocol metadata already provide sufficient discriminative power.

Despite these challenges, a small number of studies have explored payload-based traffic generation by operating directly on application-layer byte sequences~\cite{cheng2019pac, guo2021combating, wang2020packetcgan}. In these works, payloads are typically extracted after removing protocol headers and converted into fixed-length byte vectors through truncation or padding, enabling consistent representation across packets. This allows models to preserve fine-grained application semantics embedded in the payload, while still remaining constrained by practical considerations such as limited payload visibility in encrypted traffic, data availability, and computational overhead.

\noindent
\textbf{Data collection mechanisms:}
Packet capture tools such as Wireshark~\cite{wireshark} and Tcpdump~\cite{tcpdump} can record payload data in raw or hexadecimal format when traffic is unencrypted or captured at trusted endpoints (Fig.~\ref{fig:pcap-2data}). Passive monitoring frameworks such as Zeek~\cite{zeek}, however, typically do not retain full payloads and instead extract and log application-layer metadata, including Uniform Resource Identifiers (URIs), HTTP headers, and Multipurpose Internet Mail Extensions (MIME) types. Deep Packet Inspection (DPI) systems can collect and analyze payloads directly, most effectively in unencrypted traffic, using techniques such as regular expression matching and grammar-based parsing to identify protocols, extract features, and detect anomalies. In encrypted settings, payload visibility is limited, and only derived metadata or side-channel features are typically available for traffic analysis and generation.

\subsection{Transport, Network, and Data Link layers (L4--L2): Packet headers}

\begin{table*}[t!]
\scriptsize
    \centering
    \caption{Comparison of selected properties of packet and flow level data and their impact on  data generation process}
    \label{table:comparison}
    \begin{tabular}{|p{2cm}|p{4cm}|p{3.5cm}|p{6.5cm}|}
        \hline
        \textbf{Property}&\textbf{ Packet level }& \textbf{Flow level}& \textbf{Impact on data generation}\\
        \hline
        Granularity & High granularity offer more details of packet data e.g., packet header, payload. & Low granularity as only a summary of the packet stream is given. & Generation complexity depends on granularity; packet-level data with higher variability often requires more complex models. \\
        \hline
        Amount of data & High amount of data as each packet is considered & Low amount of data as a summary of each flow is recorded~\cite{hofstede2014flow} & High-volume packet-level analysis captures detailed variations but may slow model convergence~\cite{hofstede2014flow}, while lower-variability flow-level data has faster convergence with smaller datasets~\cite{lindner2023theory}\\
        \hline
        Memory usage& Requires more memory to handle large amount of data & Comparably low memory requirements~\cite{hofstede2014flow} & During model training packet level data needs more memory with large batch sizes and training time.\\
        \hline
        Processing overhead & More processing requirements to process a large number of packets & Less processing requirements with smaller datasets & For more processing requirements additional resources including CPUs, GPUs may be required.\\
        \hline
        Purity of data & With more packets, higher possibility of having noisy packet data & Impact of noisy data is often diluted by  data aggregation & Noisy inputs will be reflected in the output data reducing the quality of synthetic data~\cite{xing2024ai}\\
        \hline
    \end{tabular}
\end{table*}

The most widely used data representations in network traffic generation are derived from packet headers at the Transport, Network, and Data Link layers. This is largely because header fields are observable at many network vantage points, including routers, switches, firewalls, and monitoring appliances, even when payloads are encrypted. As a result, header-level data offers a practical and broadly accessible view of network behavior across operational environments.

Header information captures concise but expressive summaries of packet properties, such as packet size, protocol type, source and destination addresses, port numbers, and timing information. These features are sufficient for many downstream tasks, particularly in machine learning–driven traffic analysis, including traffic classification, anomaly detection, and performance modeling~\cite{madarasingha2022videotrain++, jiang2024netdiffusion, conti2015analyzing}. Consequently, most traffic generation studies operate on header-derived data rather than full payloads or physical-layer signals.

While headers at different layers expose distinct fields, the most commonly used attributes for data generation include:

\begin{enumerate}
\item \textbf{Transport layer (L4):} Transmission Control Protocol (TCP) and User Datagram Protocol (UDP) headers provide source and destination port numbers, control flags (e.g., SYN, ACK, FIN, RST), and segment-level information, which are widely used to characterize connection behavior and session dynamics~\cite{jiang2024netdiffusion,yin2022practical}.
\item \textbf{Network layer (L3):} The Internet Protocol (IP) header exposes packet length, source and destination IP addresses, time-to-live (TTL), and upper-layer protocol identifiers, enabling the modeling of routing behavior and end-to-end communication patterns~\cite{yin2022practical, xu2021stan}.
\item \textbf{Data link layer (L2):} Medium Access Control (MAC) addresses, EtherType values, and frame length provide visibility into local network topology, link-layer protocol usage, and broadcast or multicast behavior~\cite{cordero2021generating}.
\end{enumerate}

Some studies focus on data extracted from a single layer, such as frame length at the Data Link layer~\cite{sivaroopan2023synig, sivaroopan2024netdiffus}, while others combine fields across multiple layers to construct multivariate representations. For example, in~\cite{yin2022practical, jiang2024netdiffusion}, authors extract source and destination IP addresses and protocol identifiers from the Network layer together with port information from the Transport layer, enabling richer representations of communication patterns.

As shown in Fig.~\ref{fig:data-draft}, header-based information is primarily organized into two data representations: \textit{packet-level} and \textit{flow-level} data, which differ in aggregation granularity but are derived from the same observable header fields.

\subsubsection{Packet-level representation}

Packet-level data consists of raw packet captures, typically stored in \texttt{.pcap} format~\cite{lin2020using, dowoo2019pcapgan, han2019packet, jiang2024netdiffusion, cheng2019pac}. Each record in a packet trace corresponds to an individual packet and contains header fields from multiple layers, along with optional payload data (Fig.~\ref{fig:pcap-2data}). From a visibility standpoint, packet-level representations preserve fine-grained temporal dynamics and per-packet variations, making them suitable for modeling detailed protocol behavior, timing patterns, and burstiness. However, they also result in large data volumes and higher processing overhead.

\subsubsection{Flow-level representation}

Flow-level data aggregates packets into logical communication units, or flows, typically defined by a five-tuple consisting of source IP, destination IP, source port, destination port, and protocol~\cite{li2013survey}. In addition to this identifying tuple, flows often include summary statistics such as total packet count, total bytes, duration, and average throughput in both directions~\cite{citrix_appflow}. This representation is widely standardized through telemetry frameworks such as Cisco NetFlow~\cite{rfc3954}, sFlow~\cite{rfc3176}, and the IETF IP Flow Information Export (IPFIX) protocol~\cite{rfc5101}, enabling interoperable and vendor-agnostic traffic reporting across large network infrastructures. Beyond simple aggregation, flow records support the construction of statistical fingerprints of network behavior—capturing distributions of packet sizes, inter-arrival times, directional asymmetries, and temporal activity patterns—which are frequently leveraged in anomaly detection, intrusion detection, and traffic classification systems. From a visibility perspective, flow records therefore provide a compact yet analytically expressive abstraction of traffic behavior, making them suitable for large-scale monitoring, long-term analysis, and scalable traffic generation, albeit at the cost of losing fine-grained packet-level dynamics.

Packet-level and flow-level representations exhibit complementary strengths and limitations in terms of data volume, granularity, processing requirements, and noise sensitivity. Table~\ref{table:comparison} compares selected properties of these two data types and highlights their implications for traffic generation. For consistency, we assume that both representations are derived from the same underlying packet capture.

\noindent
\textbf{Data collection mechanisms:} A wide range of tools are used to collect packet header and flow data across different layers of the network stack. Packet capture tools such as Wireshark~\cite{wireshark} and Tcpdump~\cite{tcpdump} provide multi-layer visibility by extracting header information, including IP addresses at the Network layer and port details at the Transport layer. At the Transport layer, network firewalls (e.g., Cisco ASA) are commonly deployed to monitor, filter, and log connection-level activity. At the Network layer, router and switch monitoring platforms such as ManageEngine OpManager~\cite{manageengineOpManager} are widely used to collect traffic statistics, perform port-level analysis, and detect device failures. At the Data Link layer, hardware-based network Test Access Points (TAPs)~\cite{svoboda2015network} and switch-based port mirroring techniques are employed to duplicate traffic from specific links or ports, enabling passive header-level observation without interfering with traffic flow.

\subsection{Physical layer (L1): Signal-level representations}

At the Physical layer, network traffic is represented as electrical, optical, or radio-frequency (RF) signals transmitted over wired or wireless media. Unlike higher-layer representations, physical-layer data consists of waveform-level measurements that encode information through modulation, amplitude, phase, and frequency variations. These signals do not directly expose packet headers or application semantics, as higher-layer protocol structures are reconstructed only after signal demodulation and decoding. Furthermore, physical-layer signals exhibit high variability due to channel effects such as noise, interference, attenuation, and multipath propagation.

From a data visibility perspective, physical-layer representations are accessible only at specialized observation points, such as hardware taps on wired links or radio receivers in wireless environments. As a result, the collected data typically consists of raw time-domain signal samples, frequency-domain spectral representations, or derived measurements reflecting transmission and channel characteristics.

Several prior works have explored synthetic generation of such signal-level representations across different wireless technologies~\cite{yang2019generative, rf_diffusion, erol2019gan, RadioSpectrumGAN, shi2019generative, wang2025generative}. These studies operate directly on RF waveform data or their spectral representations to reproduce signal characteristics observed in Wi-Fi, radar, cellular, and RFID systems. In addition, higher-frequency signals such as millimeter-wave (mmWave) and Terahertz communications have been considered in~\cite{balevi2021wideband}. Beyond waveform generation, related physical-layer attributes such as Channel State Information (CSI), which characterizes signal propagation and channel effects, have also been synthetically generated to emulate realistic transmission environments~\cite{11059502}. Furthermore, works such as~\cite{o2018physical, o2019approximating} consider generation at the modulation and coding level using physical-layer bit sequences as input representations.

\noindent
\textbf{Data collection devices:}
At the physical layer, raw signal data is collected using specialized hardware capable of capturing transmissions prior to higher-layer decoding. In wired networks, Network Test Access Points (TAPs) provide non-intrusive access to electrical or optical signals directly from communication links. In wireless environments, Software Defined Radios (SDRs) such as RTL-SDR and HackRF~\cite{rtlsdr,hackrf} serve as primary data collection tools, enabling flexible capture of RF signals across a wide range of wireless technologies, including Wi-Fi, Bluetooth, ZigBee, and cellular systems. Additionally, wireless interfaces operating in monitor mode, supported by tools such as Aircrack-ng~\cite{aircrackng}, enable the collection of low-level transmission data and signal metadata associated with wireless communication.

\subsection{Contextual and conditional attributes across layers}


Certain factors influencing network traffic, such as user behavior, application type, available bandwidth, time of day, and connection medium, are not directly observable at any single OSI layer. Instead, these attributes reflect external context or operational conditions that shape traffic patterns across multiple layers. For example, the application in use (e.g., video streaming versus web browsing) is defined at the Application layer but manifests indirectly through packet sizes, flow durations, and throughput at lower layers. Similarly, the connection medium (e.g., fiber optics versus wireless) constrains available bandwidth and latency, influencing observable characteristics at the Transport and Network layers.

From a data visibility perspective, these attributes are typically not extracted directly from packet captures or flow records. Instead, they are inferred, logged separately by network management systems, or known a priori in controlled environments. As such, they are more appropriately treated as \emph{conditional attributes} or \emph{meta-data} that guide traffic generation rather than as primary data representations.

The synthetic generation of such contextual attributes—including application selection, user activity patterns, access technology, and usage timing—is essential for enabling realistic, controlled, and scalable traffic generation frameworks. Prior work~\cite{lin2020using, yin2022practical} has demonstrated the feasibility of generating limited sets of such meta-data and using them as conditional inputs to traffic synthesis pipelines. However, these efforts are restricted to narrow attribute spaces. More broadly, systematic generation and integration of richer, multi-dimensional contextual attributes remain largely unexplored, representing an open challenge for controlled and policy-aware traffic generation.

In this section, we presented the different types of network traffic data across the protocol stack, ranging from physical-layer signal representations to higher-layer packet and flow-level abstractions. Each data type provides a distinct level of visibility into network activity, with varying structural characteristics, semantic richness, and collection mechanisms. These differences directly influence how network traffic can be modeled and synthesized. Building on this foundation, the following sections examine synthetic data generation methods, focusing on the techniques and approaches used to generate realistic network traffic data across these different representation levels.

\section{Statistical methods based generation}\label{sec:stat}

\begin{figure*}[t]
\centering
\includegraphics[width=0.9\textwidth]{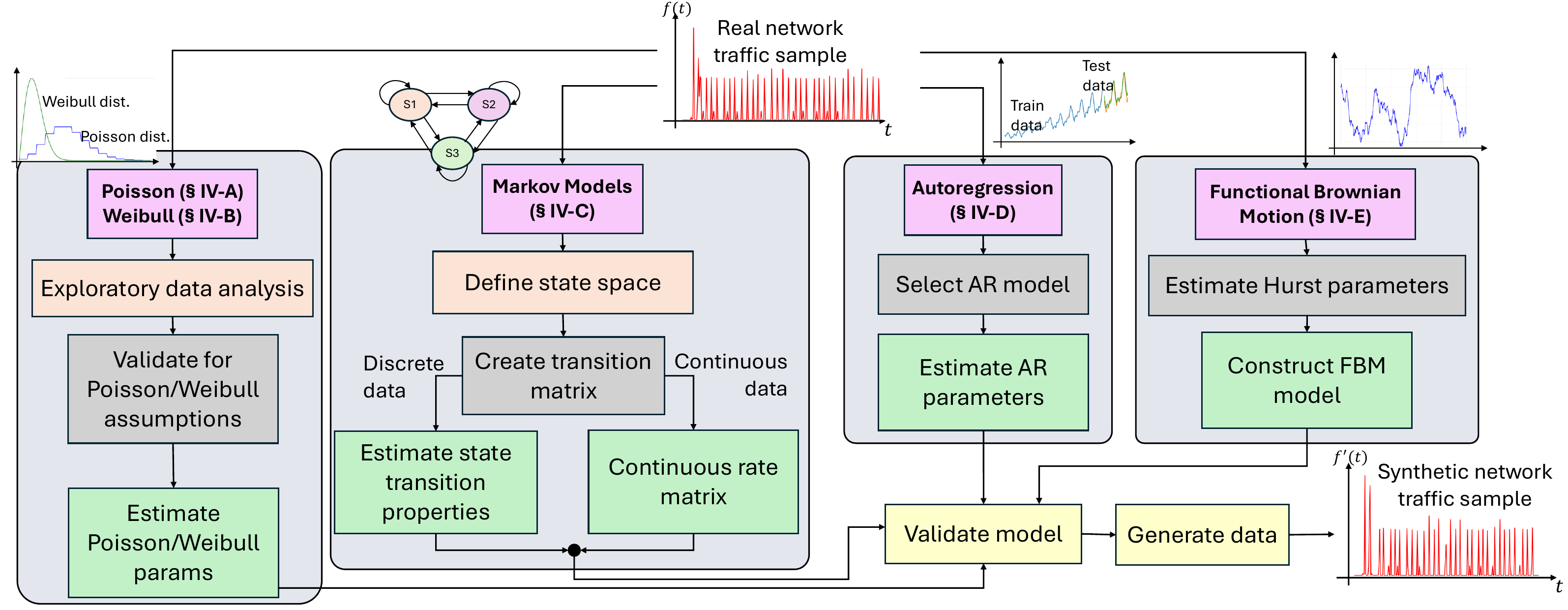}
\caption{Overall workflow of major statistical methods used.}
\label{fig:stat_overview}
\end{figure*}

In the early stages of synthetic data generation, statistical methods emerged as a foundational approach to modeling, predicting, and synthesizing network traffic. Research efforts focused on understanding the underlying characteristics of network traffic distributions to identify patterns that define network behavior. Fig.~\ref{fig:stat_overview} provides a high-level overview of the workflow for using statistical methods for data generation.

First, captured network traffic samples are processed to extract statistical properties of traffic, such as arrival rates, inter-arrival times, packet sizes, etc., and based on the characteristics of the extracted features, a suitable distribution/model (such as PP, MM, etc. that are discussed in this section) is chosen. Next, a model/distribution will be generated by estimating the corresponding parameters to fit the captured data, and the generated distribution/model will be validated against the real traffic. Once a suitable model is established, it can be used to predict future traffic patterns by changing the model parameters as necessary. These models were used to predict future traffic behavior and inform network planning and performance analysis. Additionally, synthetic traffic traces could be generated based on these models to simulate various scenarios and evaluate network performance.

Next, we present the literature found for synthetic data generation using statistical methods, categorized by each statistical model type. Initially, the fundamental principles of each model are presented, followed by an analysis of how these models have been applied to network traffic generation across various application scenarios. Table~\ref{table:stat_summary_updated} provides a summary of these studies, detailing the application scenario, the statistical model used and its adaptations, the type of data analyzed, and the availability of associated artifacts.

\begin{table*}[t!]
\scriptsize
    \centering
    \caption{Summary of works leveraging statistical methods with pros and cons.}
    \label{table:stat_summary_updated}
    \begin{tabular}{|P{0.7cm}|P{2.7cm}|P{2.5cm}|P{1.5cm}|P{2.3cm}|P{2.5cm}|P{2.8cm}|}
    \hline
    \multirow{2}{*}{\textbf{Work}} & \multirow{2}{*}{\textbf{Application scenario}} & \multirow{2}{*}{\textbf{Method(s) used}} & 
    \multicolumn{2}{|c|}{\textbf{Data}} & \multirow{2}{*}{\textbf{Pros}} & \multirow{2}{*}{\textbf{Cons}} \\
    \cline{4-5}
    &&& OSI layer & Format/Type & & \\
    \hline
    \cite{paxson1995wide} & Wide area traffic & PP & Transport & Session/Flow/Packet & Simple and tractable & Misses LRD \\
    \hline
    \cite{ns-3} & IP traffic & PPBP & Network & Flow & Captures LRD & Assumes fixed burst pattern \\
    \hline
    \cite{PoissonShotNoise_TrafficMatrix} & Data center traffic generation & PSN & Network & Flow & Models realistic DCN flows & Less suited for non-DCN traffic \\
    \hline
    \cite{yi2017modeling} & D2D mmWave modeling & Poisson Cluster Process & Physical & mmWave & Captures spatial clustering & Assumes idealized distribution \\
    \hline
    \cite{li2020modeling} & Downlink D-MIMO modeling & Poisson point process & Physical & mmWave & Models RAU/UE randomness & Ignores temporal patterns \\
    \hline
    \cite{arfeen2019role} & Access/core Internet traffic & WD & Transport/ Network & Session/ Flow/ Packet & Fits heavy/light tails & Misses correlation structures \\
    \hline
    \cite{weibull_2009} & WLAN TCP/IP modeling & Piecewise Weibull + Cascade & Transport/ Network & Flow/Packet & Captures flow inter-arrivals & Can't model intra-flow arrivals \\
    \hline
    \cite{weibull_2013} & Internet traffic modeling & WD & Transport/ Network & Session/ Flow/ Packet & Links tiers via parameters & Needs tier-specific tuning \\
    \hline
    \cite{pei2010using} & Indoor positioning & WD & Physical & RSSI & Fits RSSI well & Needs precise parameters \\
    \hline
    \cite{MUSCARIELLO20051835} & Internet traffic modeling & 5-param Hierarchical MMPP & Transport/ Network & Session/ Flow/ Packet & Captures key traffic metrics & Lacks full LRD modeling \\
    \hline
    \cite{markov_2013} & Oscillatory traffic modeling & Nested HMM (var. duration) & Transport/ Network & Arrival rate process & Matches oscillations & Higher complexity \\
    \hline
    \cite{markov_2008} & Packet-level modeling & HMMs & Application & Packet & Predicts from little data & Short-term only \\
    \hline
    \cite{garcia2015modelling} & Malicious behavior detection & Markov Chains & Transport & Flow & Good for botnet C\&C & Botnet-specific behavior \\
    \hline
    \cite{kamthe2013improving} & Wireless link simulation & HMM + MMB & Physical & Wireless & Captures multi-scale dynamics & Needs rich measurements \\
    \hline
    \cite{farima} & Traffic prediction & ARFIMA & Network & Packet & Models SRD+LRD & Fitting is complex \\
    \hline
    \cite{ARIMA-GARCH} & Traffic prediction & ARIMA/GARCH & Transport & Packet & Captures SRD, LRD, multifractal & Many parameters \\
    \hline
    \cite{pashko2018accuracy} & Self-similar traffic modeling & FBM & Not specified & Not specified & Natural self-similarity & Weak short-term modeling \\
    \hline
    \cite{10.1145/3618257.3624808} & 5G control-plane modeling & HSM & Application & Control plane events & Captures bursty deps & Complex to deploy \\
    \hline
    \cite{cieslak2006combating} & IDS imbalance handling & KNN & Not specified & Not specified & Aids minority detection & Risk of overfitting \\
    \hline
    \cite{vu2016learning} & Encrypted traffic classification & KNN & Network & Flow & Combines sampling methods & Dependent on sampling quality \\
    \hline
    \end{tabular}
\end{table*}

\subsection{Poisson process (PP)}

This is a mathematical model used to describe a sequence of random events that occur independently and at a constant average rate over time or space. It is widely used in fields like queuing theory, telecommunications, and physics to model random phenomena such as arrivals, failures, or requests. The key characteristics of a poison process includes independence (occurrences of events are independent of each other) and stationarity (the probability of an event occurring in a given interval depends only on the length of the interval, not its position in time or space). This process can be defined by:
\begin{itemize}
    \item A rate parameter \( \lambda > 0 \), which represents the average number of events per unit time or space.
    \item The number of events \( N(t) \) in a time interval \( [0, t] \), which follows a Poisson distribution as follows where \( \lambda t \) is the expected number of events in the interval \( [0, t] \): \[
  P(N(t) = k) = \frac{(\lambda t)^k e^{-\lambda t}}{k!}, \quad k = 0, 1, 2, \ldots
  \]
\end{itemize}

    Early research in modeling network traffic and synthesizing focused on using the PP to accurately model the dynamics of network traffic. A major factor for utilizing the PP, rooted in the principles of statistical independence, proved effective for certain scenarios~{\cite{NGN_Poisson}}. Research was done to explore the limitations of Poisson modeling and identified scenarios where it failed to accurately represent the observed traffic patterns~{\cite{paxson1995wide}}. In the early 1990s, two groundbreaking papers revealed that both WAN and LAN traffic exhibit Long Range Dependence (LRD) and self-similarity~{\cite{10.1145/167954.166255}, \cite{paxson1995wide}}. In addition, the authors of~{\cite{paxson1995wide}} show that PPs are valid only for modeling the arrival of user sessions (Telecommunication Network [TELNET] connections, File Transfer Protocol [FTP] control connections), but they perform poorly for other WAN arrival processes. Therefore, it became evident that the PP was inadequate for this task. Recognizing the limitations of the PP, researchers initially explored modified approaches based on the PP to capture LRD and self-similarity properties. 

    One such alternative model used is the Poisson Pareto Burst Process (PPBP). Ammar \textit{et al.}~{\cite{ns-3}} introduced a new tool for generating realistic internet traffic named NS-3, an open-source discrete-event network simulator for internet systems. By utilizing PPBP, the paper focused on capturing the LRD characteristics observed in real-world traffic. The traffic model was validated using an autocorrelation function plot that revealed the LRD property of real-world internet traffic. 

     The limitations of the PP were more evident when modeling high-speed network traffic, particularly real-time data traffic modeling for next-generation networks. For example, Liji \textit{et al.}~{\cite{NGN_Poisson}} demonstrated that the Stationary Poison Increment Process can only model Short Range Dependence (SRD) but not LRD. To address this limitation, the authors proposed using second-order self-similarity models, such as fractional Gaussian noise and Autoregressive Fractionally Integrated Moving Average (ARFIMA) processes, as a more appropriate approach.  In the meantime, researchers also explored modeling data center network traffic using PPs. To  simulate traffic in data center environments, the generation of flow-level network traffic matrices based on the Poisson Shot-Noise (PSN) model  is proposed in~{\cite{PoissonShotNoise_TrafficMatrix}}. By incorporating factors such as flow arrival rates, intra-rack traffic ratios, flow sizes and durations, the PSN process offers a more accurate representation of traffic patterns in data centers.

    Researchers have  investigated modeling signals or metrics of physical layer. In ~\cite{yi2017modeling}, the authors study the performance of mmWave communications in clustered device-to-device (D2D) networks, where D2D transceivers are spatially modeled using a Poisson Cluster Process. Each cluster features multi-antenna devices, and mmWave links are used by active D2D transmitters to serve nearby receivers. The paper proposes and analyzes three user association strategies: uniformly distributed, nearest, and closest line-of-sight (LOS) D2D transmitter models. In ~\cite{li2020modeling}, the authors analyze a downlink distributed millimeter wave massive Multiple-Input Multiple-Output (MIMO [D-MIMO]) system where radio access units (RAUs) and user equipments (UEs) are modeled as Poisson point processes. A hybrid precoding scheme based on antenna array response vectors is applied, and the impact of the RAU-to-UE ratio on system performance is examined. Two RAU allocation strategies—Distance-based and Signal-to-Interference-plus-Noise Ratio (SINR)-based—are proposed, and a lower bound on the asymptotic average spectral efficiency is derived for the Distance-based approach.

    \textbf{Limitations}: Taken together, the studies discussed in this subsection indicate that the core limitation of Poisson-based traffic models lies in their structural assumptions. The independence and stationary increment properties inherent to PPs conflict with empirical observations of long-range dependence and self-similarity reported for WAN and LAN traffic \cite{10.1145/167954.166255,paxson1995wide}. While PP-derived extensions such as PPBP and PSN partially mitigate these issues by introducing burst-level dynamics or flow aggregation \cite{ns-3,PoissonShotNoise_TrafficMatrix}, they still rely on Poisson arrivals at a fundamental level. Consequently, these models remain constrained in their ability to represent correlated behavior across multiple time scales, particularly outside the specific scenarios for which they are designed.

\textbf{Summary:} In summary, PPs form the conceptual baseline for statistical network traffic generation and have played a crucial role in shaping early modeling efforts. Their simplicity enables clear analytical insights and supports extensions into spatial and flow-level modeling at higher layers \cite{yi2017modeling,li2020modeling}. However, the empirical evidence presented in this subsection consistently shows that realistic traffic behavior requires mechanisms beyond memoryless arrivals. As a result, PP-based models are best viewed as building blocks or components within more expressive frameworks rather than complete solutions for modern traffic synthesis.

\subsection{Weibull distribution (WD)} 

As discussed earlier, the limitations of PPs for modeling network traffic led to exploring other distributions.
Overcoming the limitations of PPs, WD model was proposed mainly due to its flexibility to model both heavy and non-heavy tailed distributions~\cite{arfeen2019role}. 
The WD is a continuous probability distribution where the probability density function is as follows:
\[
f(x) = \frac{\gamma}{\alpha} \left(\frac{x - \mu}{\alpha}\right)^{\gamma - 1} \exp\left(-\left(\frac{x - \mu}{\alpha}\right)^\gamma\right) \]
Here, $x \geq \mu $ and $\gamma, \alpha > 0$ where $\gamma$ is the shape parameter, $\mu$ is the location parameter and $\alpha$ is the scale parameter. The case where $\mu = 0$ and $\alpha = 1$ is called the standard WD, while the case where $\mu = 0$ is called the 2-parameter WD.
Initial studies focused on the application of WDs to wireless TCP/IP traffic. Kullback et al.~{\cite{weibull_2009}},  highlighted the inadequacy of conventional distributions in modeling the marginal distribution of wireless TCP flow inter-arrival times. The authors proposed a piecewise WD and a canonical multinomial cascade model to capture the salient characteristics of wireless TCP flow inter-arrivals. This model achieved an acceptable accuracy for predicting packet loss rates. However, it states that while wireless TCP/IP can be partially modeled, particularly the inter-arrival times of new flows and the number of packets per flow, the arrival process of packets within a flow is not tractable due to correlations, periodicity, and inconsistency. 

Further research was done in modeling internet traffic across various network tiers utilizing the WD. Arfeen et al~{\cite{weibull_2013}} focused on how the two-parameter WD can be used in internet traffic modeling. It demonstrates how the WD captures the transformation of inter-arrival process of all structural components of internet traffic (packets, flows, and sessions) as traffic moves from access to the core network. Furthermore, the authors show that given a suitable scale parameter (media or tier specific) the Weibull shape parameter can be used to zoom in from session to flow and to packet-level inter-arrivals. 

Later in 2019, A. Arfeen et al.~\cite{arfeen2019role} provide the analytical justification for why the traffic converges to WD as traffic moves from access to core links. This convergence is attributed to the WD's flexibility in capturing the stochastic properties of internet traffic at different levels (packets, flows, sessions). The study presents a comprehensive analysis of all structural components of internet traffic, linking fractal renewal processes to the modeling of correlated traffic. Additionally, the research validates Weibull-based count and interval traffic data modeling using both data fitness and queuing analysis. 

In ~\cite{pei2010using}, the authors propose a fingerprinting-based indoor positioning method that models bluetooth signal using a WD to address the limitations of sparse Received Signal Strength Indicator (RSSI) during the training phase. By estimating the shape and scale parameters, the approach captures the full RSSI domain more effectively than conventional occurrence-based methods. For the positioning phase, a histogram maximum likelihood estimator based on bayesian theory is employed to improve localization accuracy and reliability.

\textbf{Limitations:} Although Weibull-based models significantly improve marginal distribution fitting compared to Poisson models, their limitations emerge when moving beyond distributional accuracy. As shown in \cite{weibull_2009}, WDs struggle to represent packet arrivals within flows due to correlation and periodicity effects. Moreover, while \cite{weibull_2013} demonstrates that Weibull parameters can link traffic behavior across network tiers, this linkage depends on tier-specific scale parameters, reducing model universality. These observations suggest that WDs are descriptive rather than generative in a temporal sense, capturing what traffic “looks like” statistically but not how it evolves dynamically.

\textbf{Summary:} Overall, WDs occupy an important middle ground in statistical traffic modeling. The works discussed demonstrate their ability to flexibly capture inter-arrival characteristics across packets, flows, and sessions and to explain how traffic distributions transform from access to core networks \cite{weibull_2013,arfeen2019role}. While WDs do not explicitly model temporal dependence, they provide strong empirical grounding for understanding traffic structure and serve as effective components for hybrid or hierarchical modeling approaches.

\subsection{Markov Model(MM)}

    This is a mathematical framework for modeling systems that undergo transitions from one state to another in a probabilistic manner based on the Markov property. Markov property states that the future state of the system depends only on the current state and not on the sequence of events that preceded it. 
    In contrast to the previous models, MMs offer a complementary approach, particularly capturing the temporal dependencies and oscillatory behavior often observed in network traffic. Despite the inability to achieve true LRD or self-similarity, MMs can be used to model internet traffic due to the possibility of exploiting analytical techniques to predict network performance, which according to~\cite{MUSCARIELLO20051835} is important as it is the ultimate goal when adopting models to either study existing networks or design new ones.

    Most traffic models proposed to capture LRD before Muscariello et al.~\cite{MUSCARIELLO20051835} had limited practical use in network design due to complex mathematical structures. The authors argue that long-term correlations beyond a threshold may not significantly affect performance, suggesting simpler models like Markov-Modulated PP (MMPP) are often sufficient. They introduce a 5-parameter hierarchical MMPP to model traffic across sessions, flows, and packets. Three parameters map directly to average traffic metrics (flow arrival rate, packets per flow, and packet arrival rate), while the other two (flows per session and session arrival rate) define sessions and control the Hurst parameter. The resulting synthetic traffic shows queueing behavior consistent with real-world internet traffic.

    In 2012, Y.Xie et al.~\cite{markov_2013} proposed Nested Hidden MMs (NHMMs) with variable state-duration to model oscillatory network traffic. The model synthesizes traces that preserve temporal oscillations, self-similarity, statistical properties, queuing behavior, and multiscale energy distribution. Unlike Hidden Markov Models (HMMs) and MMPPs, which face overfitting and high complexity with more states, NHMMs use a hierarchical structure of two nested chains: the first controls overall oscillations, and the second models local fluctuations. Results show NHMMs outperform empirical, multifractal, HMM, and MMPP methods in capturing traffic dynamics.
    
    The authors in {\cite{markov_2008}} utilize HMMs to model inter packet time and packet size for Internet traffic from multiple application-layer protocols. The authors evaluate the model's ability to learn, generate, and predict realistic packet-level traffic patterns through automated analysis of empirical traffic traces, focusing on the marginal distributions, auto-covariance, and cross-covariance of inter-packet time and packet size. The work shows that the proposed model is capable of predicting the short-term future behavior of the analyzed traffic based on limited monitored traffic. Unlike previous work, which primarily focused on HTTP, this study extends the application of HMMs to other application-layer protocols such as Simple Mail Transfer Protocol (SMTP).

    Garcia et al.~\cite{garcia2015modelling} applied Markov chains to detect malicious traffic by modeling botnet Command and Control (C\&C) behavior. Using flows defined by size, duration, and periodicity, they built chains of states for C\&C channels, aggregated by 4-tuples (source/destination IP, port, protocol). Each chain was stored as MM, and detection involved checking if new flow chains matched pre-trained models. Their method showe effectiveness in distinguishing botnet traffic despite variability across C\&C channels.

    In ~\cite{kamthe2013improving}, the authors present M\&M model, a multilevel framework that captures the complex short- and long-term dynamics of 802.15.4 wireless links using HMMs and Mixtures of Multivariate Bernoullis (MMBs). This method models link behavior across time scales and generates synthetic traces that  match real data in terms of packet reception rate statistics, run-length distributions, and conditional delivery patterns.

\textbf{Limitations:} The literature reviewed in this subsection highlights a fundamental trade-off inherent to Markov-based traffic models. While increasing state complexity enables richer temporal dynamics, it also raises concerns regarding parameter estimation, scalability, and practical deployment \cite{MUSCARIELLO20051835,markov_2013}. Moreover, even advanced variants such as hierarchical MMPPs and nested HMMs remain bounded by finite memory, limiting their ability to reproduce true long-range dependence. These constraints indicate that Markov models approximate correlation effects rather than fully capturing them.

\textbf{Summary:} In summary, MMs provide a structured and analytically tractable framework for modeling temporal behavior in network traffic. The studies discussed demonstrate their effectiveness in capturing oscillatory patterns, protocol-level dynamics, and queueing behavior \cite{MUSCARIELLO20051835,markov_2013}. Although they do not achieve full self-similarity, their balance between expressiveness and interpretability makes them particularly valuable for performance analysis, anomaly detection, and scenario-driven traffic synthesis.


\subsection{Autoregression (AR)}

An AR model is a type of time series model where the current value of a time series is expressed as a linear combination of its previous values (lags) and a random error term. An autoregressive model of order \( p \), denoted as AR(p), is defined as:
\[
X_t = c + \phi_1 X_{t-1} + \phi_2 X_{t-2} + \cdots + \phi_p X_{t-p} + \epsilon_t
\]
where, \( X_t \) is the value of the time series at time \( t \), \( c \) is a constant term, \( \phi_1, \phi_2, \ldots, \phi_p \) are the autoregressive coefficients, \( p \) is the number of previous time steps (lags) considered in the model, and \( \epsilon_t \) is a random error term (white noise) with mean \( 0 \) and constant variance \( \sigma^2 \).
These models are widely used in time series analysis. AR Integrated Moving Average (ARIMA)  models extend the AR model by considering the moving average and non-stationarity. ARFIMA extends ARIMA by introducing fractional differencing to account for long-memory dependence.

Shu et al.~\cite{farima} investigated ARFIMA for accurate traffic prediction, emphasizing the need to capture both SRD and LRD. They note that ARFIMA, Markov modulated processes, Transform-Expand-Sample (TES), and scene-based models can all capture SRD and LRD, but the complexity of ARFIMA fitting has limited adoption. To address this, they propose a simplified fitting procedure, observing that the order (p, q) remains stable over long periods, enabling more efficient fitting. They further introduce an adapted prediction method that provides an upper probability bound for forecasts. Their results show that ARFIMA achieves higher accuracy than AR models in h-step forecasting while requiring fewer parameters.

ARIMA/GARCH~\cite{ARIMA-GARCH} is a non-linear time series model combining linear ARIMA with Generalized Autoregressive Conditional Heteroskedasticity (GARCH) to predict network traffic. Accurate prediction requires capturing SRD, LRD, large-scale self-similarity, and small-scale multifractality. ARIMA captures only SRD, ARFIMA captures SRD and LRD but not multifractal behavior, while Multifractal Wavelet Models capture multifractality but fail in prediction. Network traffic is also non-stationary and non-linear, motivating ARIMA/GARCH, where the GARCH component models dynamic, time-varying variance to capture burstiness. The authors propose a parameter estimation procedure and an adaptive prediction scheme, demonstrating superior performance over ARFIMA in forecasting network traffic.


\textbf{Limitations: } The AR-based models reviewed reveal that improved memory modeling comes at the cost of increased complexity. As noted in \cite{farima}, ARFIMA’s ability to capture both SRD and LRD is offset by challenging parameter estimation, motivating simplified fitting strategies. Similarly, ARIMA/GARCH models introduce additional parameters to model non-stationarity and burstiness \cite{ARIMA-GARCH}, increasing computational overhead. These characteristics limit the practicality of AR-based models for large-scale synthetic trace generation, especially when real-time adaptability is required.

\textbf{Summary:} Overall, autoregressive models offer strong predictive capabilities for network traffic analysis. The progression from AR to ARFIMA and ARIMA/GARCH reflects a shift toward capturing increasingly complex dependence structures \cite{farima,ARIMA-GARCH}. As evidenced in the reviewed studies, their primary strength lies in forecasting and short-term traffic prediction rather than full traffic synthesis, positioning them as complementary tools within broader modeling pipelines.

\subsection{Fractional Brownian Motion (FBM)}

FBM is defined as a centered Gaussian process \( B_H(t) \), \( t \geq 0 \) with the covariance function,
\[\mathbb{E}[B_H(t_1)B_H(t_2)] = \frac{1}{2} \left( t_1^{2H} + t_2^{2H} - |t_1 - t_2|^{2H} \right)\]
Here, $H \in (0, 1)$ is the Hurst parameter or the Hurst index.
FBM has been considered a suitable choice for modeling network traffic due to its inherent self-similarity, a property that mirrors modern network traffic. The authors in \cite{pashko2018accuracy} study the accuracy of simulation for network traffic using FBM. The authors describe simulation methods for FBM based on its spectral representation and the property of stationary increments. The indicated methods were used for the simulation of self-similar traffic and the loading process of telecommunication networks. The simulated FBM is then used as an input process to a linear system. The goal is to construct a model that accurately approximates FBM while also considering the system's response. While the Gaussian nature of FBM aids in analyzing queuing behavior, its limitations in capturing short-term correlations and rich scaling properties of real traffic have led researchers to explore multifractal models~\cite{MUSCARIELLO20051835}. However, these models were also difficult to manage due to their analytical complexity.

\textbf{Limitations:} While FBM provides a mathematically elegant representation of self-similar traffic, its limitations become apparent when modeling realistic network behavior. As discussed in \cite{MUSCARIELLO20051835}, FBM fails to capture short-term correlations and richer scaling behavior observed in empirical traffic traces. Although multifractal models address these deficiencies, their analytical complexity reduces their usability in practical modeling and simulation tasks.

\textbf{Summary:} In summary, FBM serves as a foundational reference model for understanding long-range dependence in network traffic. Its use in simulation and system-response analysis demonstrates its value for theoretical evaluation \cite{pashko2018accuracy}. However, due to its limited expressiveness at fine time scales, FBM is best regarded as a conceptual benchmark rather than a standalone solution for comprehensive traffic generation.

\subsection{Other}

\noindent
\subsubsection{Hierarchical State-Machine-Based (HSM) Models}~These have been introduced as a more accurate approach to modeling control-plane traffic in cellular networks (e.g., 4G LTE and 5G). Traditional models, such as PPs, have proven inadequate due to their inability to capture the bursty nature and long-tailed distributions observed in control-plane traffic. J. Meng et al.~\cite{10.1145/3618257.3624808} demonstrated that a two-level HSM structure can effectively model control-plane traffic.
The authors show that this structure enables the capture of event dependencies among different types of control-plane events. Furthermore, by using a Semi-Markov process, the model can accurately represent the time spent by a User Equipment in each state, overcoming the limitations of the MM. Additionally, an adaptive clustering scheme is included to account for variations in traffic behavior based on factors such as device type and time of day. The proposed model is experimentally validated by comparing the synthetic traffic generated by the model against real-world traffic traces.


\textbf{Limitations:} Despite their demonstrated accuracy, HSM models introduce substantial modeling and deployment complexity. As shown in \cite{10.1145/3618257.3624808}, defining hierarchical states, adaptive clusters, and semi-Markov timing requires detailed domain knowledge and extensive data preprocessing. This specialization limits the generalizability of HSMs beyond control-plane traffic scenarios.

\textbf{Summary:} In literature, HSM models represent a targeted solution for control-plane traffic generation in cellular networks. By explicitly modeling event dependencies and state durations, they address limitations of memoryless statistical models \cite{10.1145/3618257.3624808}. Their effectiveness underscores the importance of protocol-aware modeling for control-plane analysis.

\noindent
\subsubsection{K-Nearest Neighbors (KNN)}~ Classifying network traffic can be a challenging task. Dataset imbalances, where categories are not evenly represented, can lead to biased classification results, particularly when using techniques that favor majority classes. \cite{cieslak2006combating}. Popular methods to balance the class distribution include random undersampling and oversampling techniques. However, these methods have notable drawbacks. Random undersampling can inadvertently discard valuable data points, while oversampling by replication may introduce bias and increase the risk of overfitting. \cite{cieslak2006combating, vu2016learning}. By considering the above issues, instead of over-sampling by replication, \cite{chawla2002smote} introduced the Synthetic Minority Over-sampling Technique (SMOTE) to over-sample by using synthetic data. Synthetic samples are generated by operating on the feature space of minority samples and their KNN. 

 With the goal of improving the true positive rate of intrusions without significantly increasing the false positives on a highly imbalanced intrusion dataset, D. A. Cieslak et al.~\cite{cieslak2006combating} evaluate and compare sampling methods including oversampling by replication, SMOTE and undersampling. Furthermore, the study introduces a clustering based implementation of SMOTE, cluster-SMOTE. cluster-SMOTE develops minority region approximations by applying k-means clustering to the set of minority examples. Then SMOTE is applied to each cluster separately. This process allows for focused improvements on a localization basis for the minority class and improves SMOTE's performance on imbalanced datasets.

 In a similar effort, \cite{vu2016learning} investigates the sampling techniques used for addressing imbalanced data by classifying encrypted and unencrypted traffic. Since the unencrypted traffic is much higher than the encrypted traffic, it results in a dataset imbalance. Apart from the earlier mentioned methods, the study also investigates the use of Condensed Nearest Neighbor for undersampling and SMOTE with a Support Vector Machine (SMOTE-SVM) for oversampling.  


 \textbf{Limitations:} The studies reviewed indicate that KNN and SMOTE-based approaches are inherently dependent on feature-space structure and data quality. As observed in \cite{cieslak2006combating,vu2016learning}, inappropriate sampling strategies can introduce bias or overfitting, and these methods do not model temporal traffic dynamics. Consequently, their applicability is limited to data balancing rather than traffic behavior synthesis.

\textbf{Summary:} In summary, KNN and SMOTE-based methods address a complementary problem in network traffic analysis: mitigating class imbalance. The reviewed works demonstrate their effectiveness in improving classification performance for intrusion detection and encrypted traffic analysis \cite{cieslak2006combating,vu2016learning}. These methods function as augmentation techniques that enhance downstream learning tasks rather than as generative traffic models.

\subsection{Summary}

While statistical methods provided a valuable foundation for network traffic modeling and synthesis, their limitations became increasingly apparent with the discovery of self-similarity, SRD, LRD, non-stationarity, and non-linear properties in network traffic (\cite{paxson1995wide, ARIMA-GARCH}). These characteristics, along with the growing complexity of modern networks, made it increasingly difficult to accurately model network traffic using traditional statistical approaches. Moreover, the abundance of available traffic data suggested that ML techniques, with their ability to learn complex patterns from large datasets, might be more efficient for modeling network traffic.

\section{DL based  generation}\label{sec:dl}
\begin{figure}[htbp]
\centering
\includegraphics[width=0.9\linewidth]{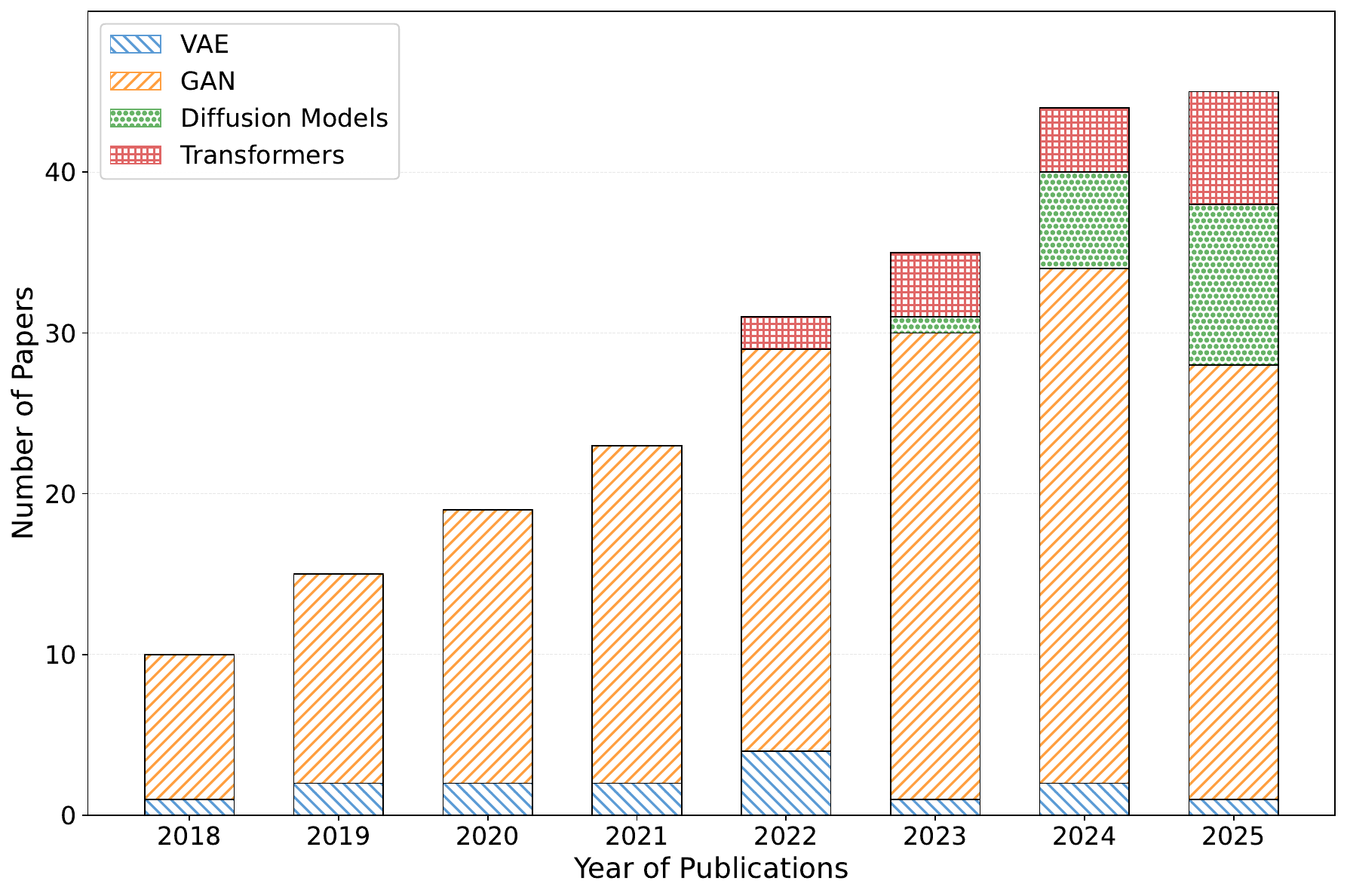}
\caption{Trends in research papers highlighting the focus on various DL models over time.}
\label{fig:paper_trend_dl}
\end{figure} 

This section reviews the main DL models used for synthetic network traffic generation: VAEs, GANs, DMs, and Transformers. Fig.~\ref{fig:paper_trend_dl} shows that GANs were the most widely used model in 2022, with 24 publications. Since then, there has been growing interest in Transformers and DMs, with publications on these models increasing from 2 in 2022 to 10 in 2024. This shift indicates a move towards more advanced models like DMs and Transformers (e.g., GPT), due to their ability to generate higher-quality data and better capture the time-based patterns in network traffic.

The remainder of this section provides a summary of the studies categorized by each model type. Initially, the fundamental architecture of each model is presented, followed by an analysis of how these models have been adapted for network traffic generation in various application scenarios. Furthermore, a brief discussion is included on the challenges associated with utilizing these models for network traffic data. In addition, similar to Section~\ref{sec:stat}, each study is summarized in a tabular format, reporting the application scenario, model and its adaptations, data type, and artifact availability.

\subsection{Variational Autoencoder (VAE)}

\begin{table*}[t!]
\scriptsize
    \centering
    \caption{Summary of VAE-based work reporting application scenario, model architecture, and the data being used.}
    \label{table:vae summary}
    \begin{tabular}{|>{\centering\arraybackslash}m{1.2cm}|>{\centering\arraybackslash}m{0.5cm}|>{\centering\arraybackslash}m{4cm}|>{\centering\arraybackslash}m{3.25cm}|>{\centering\arraybackslash}m{2.25cm}|>{\centering\arraybackslash}m{2.9cm}|>{\centering\arraybackslash}m{0.8cm}|}
    \hline
    \multirow{2}{*}{\textbf{Model}} & \multirow{2}{*}{\textbf{Work}} & \multirow{2}{*}{\textbf{Application Scenario}} & \multirow{2}{*}{\textbf{Model}} & \multicolumn{2}{|c|}{\textbf{Data}} & \textbf{Artifacts} \\ 
    \cline{5-6}
    \textbf{Variant}&&&\textbf{details}&\textbf{OSI layer}&\textbf{Format/Type}&\textbf{available}\\
    \hline

    \multirow{1}{*}{Vanilla VAE} 
    & \cite{kakkavas2021future} & Generate network traffic matrices for network management & Vanilla VAE & Data link & Traffic metrics (total bytes/5 min) & Yes \\
    \cline{2-7}
    & \cite{ha2020food} & Food sensing & Vanilla VAE & Physical & RF signals & Yes \\
    \cline{2-7}
     & \cite{liu2021fire} & Estimating downlink wireless channel for MIMO & Vanilla VAE & Physical & RF signals & No \\
    \cline{2-7}
    & \cite{yang2022rethinking} & WI-FI fall detection & Vanilla VAE & Physical & WI-FI & No \\
    \cline{2-7}
    & \cite{suroso2022deep} & Indoor localization fingerprinting & Vanilla VAE & Physical & RSSI & No \\
    \cline{2-7}
    \hline

    \multirow{2}{*}{\parbox{1.5cm}{Conditional VAE\\ (CVAE)}} 
    & \cite{aceto2024synthetic} & Attack/Intrusion detection & CVAE & Transport & Packet level & Yes \\
    \cline{2-7}
    & \cite{liu2022intrusion} & Attack/Intrusion detection & Vanilla VAE, CVAE & Transport, Network, Data link & Packet level & No \\
    \cline{2-7}
    & \cite{baur2022variational} & Channel estimation & CVAE & Physical & CSI data & No \\
    \cline{2-7}
    
    \hline

    \multirow{2}{*}{\parbox{1.5cm}{VAE with \\ Sequential Models}} 
    & \cite{xiao2018deep} & Modeling QoS & VAE + LSTM & Transport, Network & Packet level, QoS data (e.g., delay/loss) & No \\
    \cline{2-7}
    & \cite{meslet2022necstgen} & Generic network traffic generation & VAE + RNN + GMM & Transport, Data link & Packet and Flow level & Yes \\
    \hline
    \multirow{1}{*}{\parbox{1.5cm}{Misc. approaches}} 
    & \cite{bano2024variational} & Generic network traffic generation & DNN-based noise remover + LSTM + CVAE & Transport, Network, Data link & Packet level & No \\
    \cline{2-7}
    \hline

    \end{tabular}
\end{table*}

\begin{figure}[h!]
\centering
\includegraphics[width=\linewidth]{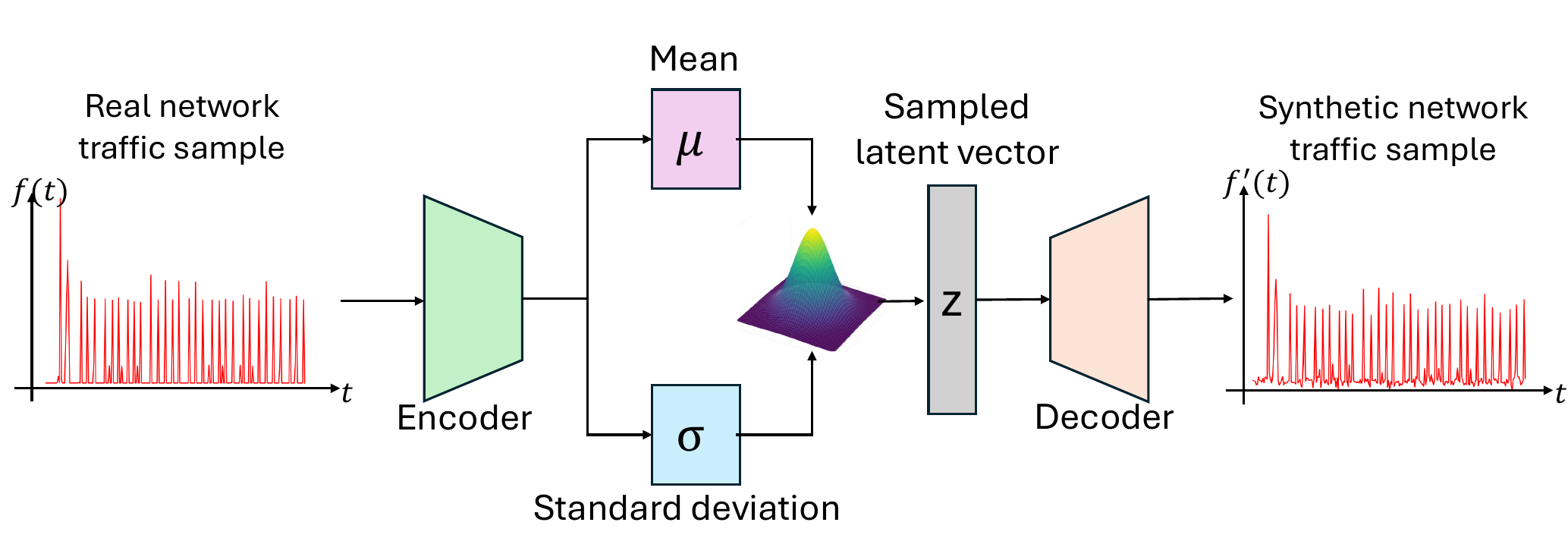}
\caption{Basic functionality of VAE}
\label{fig:VAE}
\end{figure}

Autoencoder is a neural network architecture designed to learn efficient data representations, consisting of an encoder that compresses input data into a lower-dimensional latent space and a decoder that reconstructs the original input from this compressed representation. While effective for dimensionality reduction and feature learning, traditional autoencoders lack a probabilistic framework for generating new data. The VAE, introduced in \cite{kingma2013auto} in 2013, built upon autoencoders by incorporating principles from variational inference\cite{blei2017variational}, which is shown in Fig.~\ref{fig:VAE}. We denote this basic VAE as vanilla VAE.  

Instead of encoding inputs to arbitrary points in the latent space, the VAE's encoder outputs parameters of a probability distribution, typically Gaussian, for each latent dimension. This probabilistic encoding allows for sampling from the latent space, enabling the generation of new, unseen data. The VAE's decoder then learns to reconstruct the input from these sampled latent representations. This architectural shift, combined with a modified loss function that balances reconstruction quality with the regularity of the latent space, transformed the deterministic autoencoder into a powerful generative model capable of both compression and novel data synthesis. In literature, we have observed that this basic VAE is often combined with other models to enhance synthetic output. For example, to maintain long-term relationships in network time-series data, Recurrent Neural Network (RNN) or its variations (e.g., Long short-term memory [LSTM])~\cite{xiao2018deep, meslet2022necstgen} are integrated with VAE. Another variant of VAE is Conditional VAE (CVAE) ~\cite{bano2024variational, aceto2024synthetic}, which introduces conditional labels or attributes into the model to guide the generation process. 
Table~\ref{table:vae summary} provides a summary of existing work.

\subsubsection{\textbf{Vanilla VAE}}
Kakkavas \textit{et al.}~\cite{kakkavas2021future} focuses on generating synthetic network traffic matrices (TM) to aid in network management and traffic engineering. TM generation is necessary because direct measurements of traffic are costly and difficult to obtain at scale. By using historical data to train a vanilla VAE, the model learns the underlying traffic patterns and generates new traffic matrices with similar characteristics, thus reducing the need for extensive real-time measurements. The VAE works by encoding input traffic matrices into a low-dimensional latent space, sampling from this latent space, and decoding the samples back into realistic traffic matrices. This approach helps network administrators synthesize traffic matrices that can be used for testing and simulations while also enabling the estimation of unobserved TMs from readily available link load data.

RF-EATS~\cite{ha2020food}, a noninvasive system leveraging VAE, uses passive backscatter tags and near-field coupling to sense food and liquids inside closed containers without contact or calibration. FIRE~\cite{liu2021fire}, leveraging VAE, proposes an end-to-end machine learning system for accurate downlink channel estimation in massive MIMO without requiring feedback from client devices. FIRE learns a latent representation from uplink channels to generate downlink estimates, offering an interpretable, efficient, and scalable solution for real-world deployments. In ~\cite{yang2022rethinking}, the authors present FallDar, a Wi-Fi-based DL fall detection system that addresses key challenges of environmental, motion, and user diversity. FallDar uses body speed as a robust feature, generates diverse fall scenarios with VAE, and incorporates a user identification network to enable accurate detection without needing user-specific fall data. In ~\cite{suroso2022deep}, the authors address the high cost and effort of constructing offline databases for fingerprinting-based indoor localization by leveraging VAE. Using real-world RSSI data, the VAE learns the distribution of RSSI to generate synthetic values, mitigating the challenges posed by its inherent instability. This approach aims to enhance the efficiency and reliability of DL-based indoor localization systems.

\subsubsection{\textbf{Conditional VAE (CVAE)}}

CVAE are used for addressing the class imbalance issue, particularly in applications such as IDS, where the adversarial action can be less prevalent compared to legitimate users. In~\cite{aceto2024synthetic}, the authors provide class labels as the conditional input when training CVAE for generating packet-level data. They propose an effective binning approach of the input samples to represent the distribution of characteristics. This binning process has helped the stable CVAE training process, while reducing the impact of noise samples through aggregation of feature values at the bin level. Also, it helps preserve the privacy of the original because the reverse mapping of bins and original data is infeasible.

Similarly, the work in~\cite{liu2022intrusion} uses CVAE to address the generation of network traffic for IDS by addressing the data imbalance problem in attack types, which negatively affects IDS performance. Taking the features extracted by CICFlowMeter~\cite{cicflowmeter} as the input, the VAE is employed for data augmentation, generating new synthetic samples of attack data that mimic the distribution of the original samples to increase the number of minority class samples. The CVAE, which includes class labels as the conditional input, is used to specifically generate synthetic samples for particular attack types, ensuring that each attack class is balanced in the dataset.

In ~\cite{baur2022variational}, a CVAE is proposed for data-driven channel estimation by modeling the unknown channel distribution as a conditional Gaussian parameterized by conditional moments. The authors introduce three VAE-based estimators, showing that they can approximate the minimum mean square error (MMSE) estimator, with practical variants requiring full channel knowledge only during training or not at all.

\subsubsection{\textbf{VAE with sequential models}}
To address problems such as traffic generators not being able to adapt to all types of protocols and not generating additional information like Signal to noise ratio (SNR) authors of ~\cite{meslet2022necstgen} developed an approach named Network Clustering Sequential Traffic Generation (NeCSTGen). The authors combine VAE with RNN to generate live traffic that reproduces the original behaviour at the packet, flow, and aggregate levels. Once the VAE projects the packet features into a latent space, where packets with similar features are clustered using a Gaussian Mixture Model (GMM). The RNN is then employed to ensure coherence between consecutive packets within the same flow by learning and generating transitions between these clusters, maintaining short-term dynamics and flow-level consistency. In \cite{xiao2018deep}, the authors developed LSTM-enhanced VAE  to model the QoS. The QoS inference problem is defined as inferring the exact value of QoS metrics at each moment given real-time traffic loads. The LSTM module is used to extract fine-grained traffic features as a condition to the VAE and the VAE module learns to reconstruct the QoS distribution from the traffic features.

\subsubsection{\textbf{Miscellaneous approaches}}
In ~\cite{bano2024variational}, a CVAE combined with LSTM is utilized for generating synthetic network traffic and imputing missing data within traffic streams. The data generation process involves two stages. First, a DNN model is applied as a preprocessing step to remove noisy packets and retain only legitimate traffic. Second, the legitimate traffic is used to train the CVAE, with conditional inputs (e.g., traffic attributes or labels) provided at the decoder. LSTM models are employed in both the encoder and decoder to capture temporal relationships within the traffic data. The generated data aligns with the statistical distribution of the original traffic, preserving real-world dynamics even with missing data.

\subsubsection{\textbf{Limitations in VAE based traffic generation}}
Although existing research does not directly identify limitations of VAEs specific to traffic generation, their general limitations can be linked to the task of network traffic generation. First, VAEs often produce blurry reconstructions~\cite{berger2020variational}, which may result in the inability to replicate high-frequency components present in network traffic patterns. Second, the posterior collapse in VAE~\cite{lygerakis2023cr}, where the encoder fails to effectively learn the latent space from input data, makes the decoder put greater effort into synthesizing new data. This issue affects the modeling of complex feature variations in network traffic traces, especially those with longer durations and long-term dependencies.
Third, balancing the trade-off between latent space structure and reconstruction quality~\cite{lin2019balancing, asperti2020balancing} is another challenge in VAE. For instance, over-regularization may simplify complex traffic patterns, leading to less diverse and unrealistic data. Conversely, under-regularization may overly emphasize reconstruction fidelity, resulting in unstructured latent spaces and incoherent synthetic traffic samples.

\subsubsection{\textbf{Summary}}
VAEs have established themselves as valuable tools in network traffic generation, effectively modeling complex network traffic distributions. Table \ref{table:vae summary} provides a comprehensive summary of the work done with VAEs. From vanilla VAE to its variants, such as CVAE and combinations with RNN models (e.g., LSTM), have been applied to learn both short and long-term temporal characteristics with different network applications. These applications span quality of service prediction, anomaly detection, and intrusion prevention, covering diverse data formats, including packet and flow levels, traffic metrics, and QoS data. Despite the benefits of using VAE, challenges remain, such as posterior collapse and blurry reconstructions particularly when modeling long-term feature variations of network traffic patterns.

\begin{table*}[]
\scriptsize
    \centering
    \caption{Summary of GAN-based work reporting application scenario, model architecture, and the data being used.}
    \label{table:gan summary}
    \begin{tabular}{|>{\centering\arraybackslash}m{0.8cm}|>{\centering\arraybackslash}m{0.7cm}|>{\centering\arraybackslash}m{5.25cm}|>{\centering\arraybackslash}m{3.3cm}|>{\centering\arraybackslash}m{2.25cm}|>{\centering\arraybackslash}m{1.5cm}|>{\centering\arraybackslash}m{0.8cm}|}
    \hline
    \multirow{2}{*}{\textbf{Model}} & \multirow{2}{*}{\textbf{Work}} & \multirow{2}{*}{\textbf{Application Scenario}} & \multirow{2}{*}{\textbf{Model}} & \multicolumn{2}{|c|}{\textbf{Data}} & \textbf{Artifacts} \\
    \cline{5-6}
    \textbf{Variant}&&&\textbf{details}&\textbf{OSI layer}&\textbf{Format/Type}&\textbf{available}\\
    \hline

    \multirow{6}{*}{\parbox{1.15cm}{Vanilla \\ GAN}} 
    & ~\cite{nukavarapu2022miragenet} & Generic network traffic generation & Vanilla GAN & Transport, Network & Packet level & No \\
    \cline{2-7}
    & ~\cite{cheng2019pac} & Generic network traffic generation & Vanilla GAN & Application & Payload & No \\
    \cline{2-7}
    & ~\cite{noel2024novel} & Generic network traffic generation for darknet data & Vanilla GAN & Transport, Network & Packet level & No \\
    \cline{2-7}
    & ~\cite{xiao2023distributed} & Generic network traffic generation & Vanilla GAN, FL & Transport, Network & Packet level & No \\
    \cline{2-7}
    & ~\cite{guo2021combating} & Improve ML accuracy/class imbalance & Vanilla GAN & Application & Payload & No \\
    \cline{2-7}
    & ~\cite{shahriar2020g} & Handle class imbalance, Attack/Intrusion detection & Vanilla GAN & Transport, Network & Packet level & No \\
    \hline

    \multirow{6}{*}{\parbox{1.15cm}{WGAN}}
    & ~\cite{ring2019flow} & Generic network traffic generation & WGAN & Transport & Flow level & No \\
    \cline{2-7}
    & ~\cite{fathi2020gan} & Traffic obfuscation & WGAN & Transport, Network & Flow level & No \\
    \cline{2-7}
    & ~\cite{zhang2024sa} & Traffic obfuscation & WGAN & Transport & Packet level & No \\
    \cline{2-7}
    & ~\cite{mozo2022synthetic} & Attack/Intrusion detection & WGAN & Transport & Flow level & No \\
    \cline{2-7}
    & ~\cite{kattadige2021videotrain} & Improve ML accuracy/class imbalance & WGAN & Transport & Packet level & Yes \\
    \cline{2-7}
    & ~\cite{madarasingha2022videotrain++} & Improve ML accuracy/class imbalance & WGAN & Transport & Packet level & Yes \\
    \cline{2-7}
    & \cite{sivaroopan2023synig}& Improve ML accuracy/class imbalance & WGAN & Transport, Network & Packet level& Yes\\
    \cline{2-7}
    & \cite{RadioSpectrumGAN}& Generic RF signal generation & WGAN & Physical & Raw signal  & No\\
    \cline{2-7}
    & \cite{balevi2021wideband}& Wideband channel estimation & WGAN & Physical & mmWave  & No\\
    \hline

    \multirow{4}{*}{\parbox{1.15cm}{CGAN}}
    & ~\cite{jiang2024towards} & Generic network traffic generation & CGAN & Transport, Network & Flow level & No \\
    \cline{2-7}
    & ~\cite{wang2020packetcgan} & Improve ML accuracy/class imbalance & CGAN & Application & Payload & No \\
    \cline{2-7}
    & ~\cite{kotal2024kinetgan} & Improve ML accuracy/class imbalance & CGAN & Transport, Network & Flow level & No \\
    \cline{2-7}
    & ~\cite{almasre2024create} & IoT traffic generation for cybersecurity applications & CGAN & Transport, Network & Packet \& Flow & Yes \\
    \cline{2-7}
    & ~\cite{erol2019gan} & Human activity recognition & ACGAN & Physical & RF signal & No \\
    \cline{2-7}
    & ~\cite{doshi2022over} & mmWave MIMO channel estimation & CGAN with FL & Physical & MIMO matrix & No \\
    \cline{2-7}
    & ~\cite{xu2025gansec} & Anomaly detection & CGAN & Physical & Wireless & No \\
    \hline

    \multirow{4}{*}{\parbox{1.15cm}{Seq. \\ Model Variants}} 
    & ~\cite{lin2020using} & Class imbalance / Privacy-preserved data sharing & GAN, LSTM & Transport & Flow level & Yes \\
    \cline{2-7}
    & ~\cite{yin2022practical} & Class imbalance / privacy-preserved data sharing & GAN, LSTM & Transport, Network & Packet \& Flow  & Yes \\
    \cline{2-7}
    & ~\cite{zhang2024transflowgan} & Improve ML accuracy/class imbalance & GAN, Transformer & Transport, Network & Flow level & No \\
    \cline{2-7}
    & ~\cite{shahid2020generative} & Attack/Intrusion detection & Autoencoder, GAN, LSTM & Transport & Packet level & No \\
    \cline{2-7}
    & \cite{qu2024towards}& Attack/Intrusion detection & GAN, LSTM & Transport & Packet level &   No\\
    \hline

    \multirow{6}{*}{\parbox{1.15cm}{Misc. \\ Approach}} 
    & ~\cite{anande2023generative} & Generic network traffic generation & CTGAN, Copula GAN & Network, Transport & Flow level & No \\
    \cline{2-7}
    & ~\cite{meddahi2021sip} & Generic network traffic generation for SIP protocol & DCGAN & Application & Payload & Yes \\
    \cline{2-7}
    & ~\cite{yang2024research} & Generate network traffic & Swing GAN, Autoencoders & Transport, Network & Flow level & No \\
    \cline{2-7}
    & ~\cite{kim2024network} & Simulation tool development & GAN-based tool & Transport, Network, Data-link & Packet level & No \\
    \cline{2-7}
    & \cite{dowoo2019pcapgan}& Handle data shortage in cyber-domain& Graph/Image/Sequence GAN, Makrov models & Transport, Network & Packet level &   No\\
    \cline{2-7}

    &\cite{khan2024secure}& Handle class imbalance & BEGAN & N/A & N/A & No\\
    \hline

    \end{tabular}
\end{table*}

\subsection{Generative Adversarial Networks (GAN)}

\begin{figure}[h!]
\centering
\includegraphics[width=\linewidth]{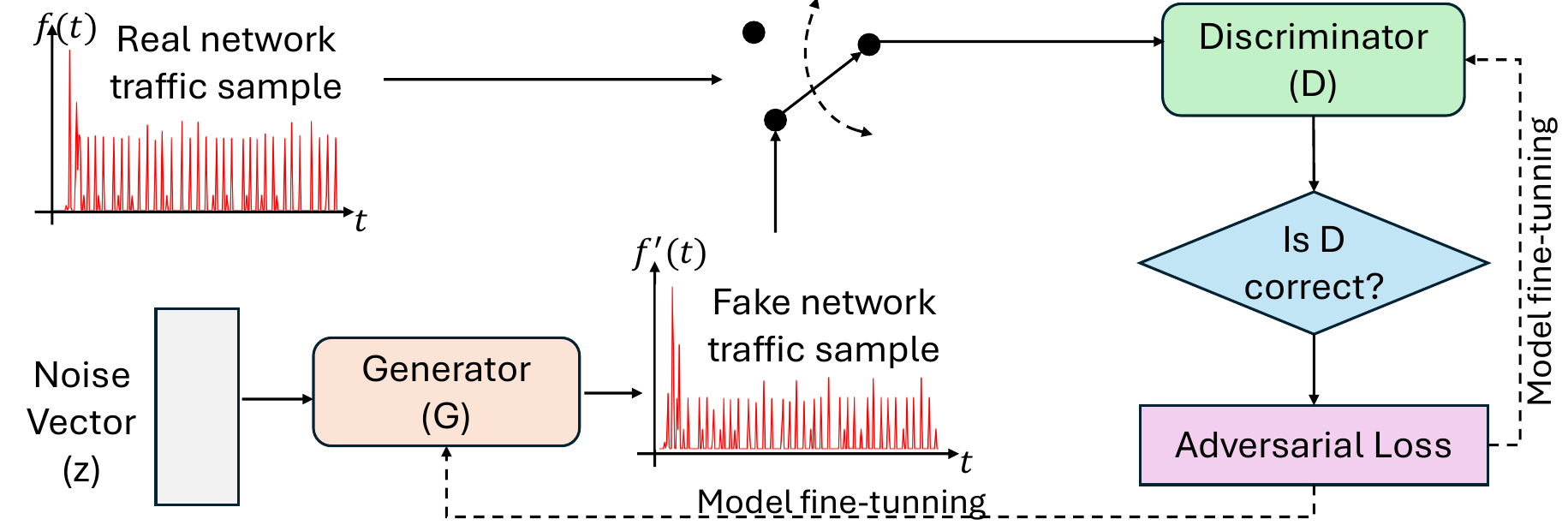}
\caption{Basic functionality of a GAN}
\label{fig:GANS}
\end{figure}

GANs, introduced by Goodfellow \textit{et al.}~\cite{goodfellow2014generative} in 2014, represented a groundbreaking approach in generative modelling. As shown in Fig.~\ref{fig:GANS}, GANs consist of two competing neural networks: a generator and a discriminator. The generator's objective is to capture the data distribution and generate realistic samples, while the discriminator's role is to differentiate between real samples from the training data and fake samples produced by the generator. The two models are trained simultaneously in an adversarial process. The generator attempts to produce data that closely resembles the real data, aiming to fool the discriminator. Meanwhile, the discriminator works to improve its accuracy in distinguishing between genuine data and the data generated by the generator. This adversarial setup results in a continuous improvement loop: as the generator becomes better at creating realistic data, the discriminator must also become more adept at distinguishing between real and fake data. The training procedure can be described as a min-max two-player game, where the generator seeks to maximize the probability the discriminator makes a mistake, while the discriminator attempts to minimize the mistakes. Although GANs were initially used for image data, researchers began applying them to time-series data including network traffic data. We identify five main GAN-based model architectures in this domain.

\subsubsection{\textbf{Vanilla GAN}} 
The basic GAN model, which comprises of only generator and discriminator modules, has shown promising results in maintaining network traffic properties, despite the sequential nature they present, in network traffic generation. The authors in~\cite{nukavarapu2022miragenet} focused on creating a protocol-agnostic framework for packet generation developing a platform named \textit{MiragePkt}. In their approach, network packets were modeled as sequences of bytes. To create these sequences they leverage packet tokenization techniques, and GANs were employed to understand the conditional distribution of bytes within these sequences. To properly understand the properties of sequence, \textit{MiragePkt} consists of 1D Convolutional Neural Network (CNN) layers. They further apply post-processing techniques to convert the output byte data from the GAN model to a hex stream.    

Generating bytes in hexadecimal form through basic GAN has been studied by Cheng \textit{et al.}\cite{cheng2019pac}. They devise an encoding scheme to convert network traffic data into matrix representations suitable for use with CNNs, which serve as the internal layers of the vanilla GAN model. This encoding technique involves transforming packet bytes (in hex format) into a matrix format (image), facilitating the integration of GANs with CNN architectures. Deviating from the conventional network traffic data, Noel \textit{et al.} \cite{noel2024novel} proposed a method for generating darknet-related network traffic, crucial for the early detection of malicious activity. The approach utilizes a vanilla GAN combined with a feature selection mechanism to pre-process and clean the data effectively. 

Advancing the usage of vanilla GAN models, Xiao et al.\cite{xiao2023distributed} proposed a federated learning(FL)-based framework for distributed GAN-based network traffic generation with self-supervised learning. The approach involves local model training at the edge, where each edge node contains multiple generators, a discriminator, and a classifier. Each generator is trained on a local dataset, while the discriminator helps improve the performance of all generators at the edge. Locally trained discriminators and generators from all edge nodes are then aggregated globally using the Federated Averaging (FedAvg) algorithm. The classifier at each edge provides pseudo-labels, enabling self-supervised training of the GAN models and facilitating autonomous training of the entire system.

A fundamental purpose of network traffic generation is to address data imbalance for ML-related applications. In~\cite{guo2021combating}, the authors utilize a basic GAN setup to address data imbalance among various protocols and applications, noting that simple oversampling of minority classes does not yield satisfactory classification accuracy. They first pre-train a classification model with the initial imbalanced dataset and then refine it using synthetic data generated by GANs, improving the performance of classification models. They leverage bytes collected at the Application layer in hexadecimal form for each packet as the data for generation, by vectorizing and removing the packet header data from network packets. 

In other network applications, vanilla GANs are utilized to generate network traffic data to enhance network IDS, particularly when class imbalance is prevalent in anomalous data. In this effort, Shahriar \textit{et al.}~\cite{shahriar2020g} apply an encoding approach to convert categorical input features into a low-dimensional continuous space. They also employ feature normalization techniques and reduce feature dimensionality using Principal Component Analysis (PCA) to support the GAN model training process.

\subsubsection{\textbf{Wasserstein GAN (WGAN)}}
WGAN is a GAN variant that incorporates the Wasserstein-1 loss (i.e., earth mover distance) into its training loss function, offering more stable training and addressing the vanishing gradient problem commonly observed in vanilla GAN models.
In generic traffic generation using WGAN, Ring et al. propose a flow-level data generation approach to address the challenge of handling both continuous and categorical data in network flows~\cite{ring2019flow}. Before training the WGAN model, they employ three techniques to convert categorical variables into a continuous domain: treating categorical attributes like IP addresses and ports as numerical values, creating binary attributes from categorical features, and using the IP2Vec~\cite{ring2017ip2vec} method to learn meaningful vector representations of categorical features. These transformed features are then used to train the WGAN model, employing the two-time-scale update rule proposed by~\cite{heusel2017gans}.

In ~\cite{fathi2020gan}, the authors leverage WGAN to obfuscate network traffic, reducing the detection capabilities of various applications. Initially, they derive statistical features for the flow level data they collected followed by normalization across the flows. Then, they use WGAN to collect network traffic data and generate flow-level statistics, followed by applying WGAN again to synthesize network packet traces based on these synthetic flow-level statistics to produce obfuscated traffic. The model has multiple WGAN modules trained on different network traffic based on applications (e.g., Google Chrome, Google Drive), enabling application-level traffic obfuscation.

To achieve traffic obfuscation through a camouflage approach, \cite{zhang2024sa} enhance the WGAN model by integrating a self-attention mechanism, enabling it to better capture feature patterns. The camouflaged features are designed so that the modified TCP packet streams, after embedding these features, meet predefined thresholds for both the Jaccard index and cosine similarity when compared to the original packet streams.

WGANs are also used in~\cite{mozo2022synthetic} for network traffic anomaly detection.  Here, WGANs are used to synthesize flow-level data aimed at detecting anomalous traffic related to cryptomining attacks, where attackers exploit computational resources without consent. To mitigate the mode collapse issue, they implemented adaptive mini-batches and introduced noise into the discriminator to prevent it from converging too quickly. This approach allowed the generator more time to improve during training. Additionally, they incorporated a multi-point single-class embedding strategy for sampling noise, utilizing uniformly distributed centroids within the noise space. 

In~\cite{kattadige2021videotrain}, features are extracted from packet-level network traffic data. During pre-processing, network packets are grouped into predefined intervals (bins), and statistical features, such as total bytes downloaded or uploaded, are derived. Percentile mapping is then applied to scale the feature values to a range of 0–100\%, reducing their variability and enhancing the training process of the WGAN model used for data generation
While this approach helps structure the data for better synthesis, it processes the original traces individually, leading to increased computational overhead and longer processing times. To mitigate this computational overhead, the same authors propose effective algorithms to control the WGAN model training in~\cite{madarasingha2022videotrain++}. This algorithm compares the fidelity between the original and synthesized data at pre-defined epoch intervals and stops the model training once the data fidelity achieves the expected levels.

Researchers have applied conversion of network traffic data into the image domain, where WGANs show great promise, after making multiple modifications to improve GANs for 1D data generation. In~\cite{sivaroopan2023synig}, the authors employ the Gramian Angular Summation Field (GASF) imaging mechanism to transform certain attributes of network traces into visual representations~\cite{wang2015imaging}. These images are utilized for network traffic generation through a simple WGAN framework. This imaging approach effectively captures the correlations between time series points, which previous studies attempt to model using more complex GAN architectures incorporating LSTMs and other networks. The authors develop a post-processing mechanism for the generated images and evaluate the model across multiple traffic types, demonstrating improved performance with the straightforward WGAN for image generation.

WGANs are used in RF signal synthesis. However, training GANs on long time-domain sequences of RF signals is challenging. To address this, ~\cite{RadioSpectrumGAN} proposed representing RF signals in the spectral domain, allowing the use of existing GAN architectures. The authors consider 2.5s long 4G LTE signal samples at 40 MSamples/sec which were transformed into 256x256 compact spectrograms, where the horizontal axis represented the time dimension over approximately 2 ms, and the vertical axis represented the frequency offset within a 40 MHz band. Each spectrogram served as input to the GAN, effectively framing the RF signal generation as a 2D problem similar to image synthesis. The authors also experimented with direct training of GANs on time series data. They found that while this approach could generate raw I/Q samples, it was significantly more challenging to capture the fine-grained features and multiscale details of LTE signals compared to the spectral domain representation. Further, in ~\cite{balevi2021wideband}, the authors propose a WGAN-based approach for estimating frequency-selective channels at high frequencies (e.g., mmWave, THz) using few pilots and operating under low SNR conditions. The generative model is first trained to capture the true channel distribution and is then used as a prior to estimate current channels by optimizing its input based on received signals.
 
\subsubsection{\textbf{Conditional GAN (CGAN)}}
CGANs extend the original GAN framework by incorporating additional information (called conditions) into both the generator and discriminator, enabling more controlled and targeted data generation. Several papers have taken this advantages in network traffic generation. For example, Jiang \textit{et al}.\cite{jiang2024towards} propose an IoT traffic generation approach for flow-level data in Non-Terrestrial Networks, providing traffic class (network application label)  as the conditional input to the CGAN. During pre-processing steps, IP2Vec is used to transform categorical variables (e.g., IP addresses) into an embedding space, which is combined with pre-processed statistical features from the flow. For the synthesized categorical data, the nearest neighbor from the real data is identified based on cosine similarity between the embedding spaces of the synthetic and real data.

Preserving class-specific details in synthetic data is essential for achieving higher classification accuracy in downstream machine learning tasks. CGANs address this by incorporating conditional inputs into the generation process. In~\cite{wang2020packetcgan}, the authors use class labels, encoded as vectors, as conditional inputs for the CGAN model. During pre-processing, packet headers are removed, and the remaining data in bytes is normalized. The resulting time series is then transformed into an image format before CGAN training. Their evaluations demonstrate that the CGAN-based approach achieves superior results compared to vanilla GAN methods.

Kotal \textit{et al.} \cite{kotal2024kinetgan} introduced a domain knowledge-guided CGAN-based approach for network traffic generation, addressing the challenge of limited real data in GAN training. Knowledge-based rules are incorporated as conditions to the CGAN in learning subtle feature variations and requirements (e.g., specific port numbers for different protocols such as port 80 for HTTP), reducing the need for additional data for CGAN training. The method leverages the concept of a unified cybersecurity ontology to develop a knowledge base tailored to network traffic data.

Almasre \textit{et al.}~\cite{almasre2024create} utilize CGANs to generate IoT network traffic, including various types of cyberattacks, to improve the analysis of security vulnerabilities. Their approach involves pre-processing steps to encode categorical features, scale numerical features, and select the most relevant attributes for generation. The CGAN model uses ground truth labels combined with an input noise vector to generate realistic and diverse network traffic. In their evaluation across various attack types (e.g., SYN Flood, UDP Flood, Reconnaissance), they demonstrate that the proposed method achieves high data fidelity, particularly in scenarios with a very limited number of network attack samples.

In ~\cite{erol2019gan}, the authors propose Auxiliary CGAN (ACGAN) that extends adversarial learning to generate synthetic radar time-frequency domain signatures for human activity recognition. This approach uses  Time-Frequency images for data synthesis. In ~\cite{doshi2022over}, the authors propose an unsupervised over-the-air (OTA) algorithm that trains CGAN with federated training setup using only noisy pilot measurements to estimate high-dimensional beamspace MIMO channels. GANSec~\cite{xu2025gansec}, a CGAN framework, is proposed for augmenting wireless time-series data to improve anomaly detection under data scarcity and class imbalance. They explore various architectures and training objectives, demonstrating that detectors trained solely on GANSec-generated data perform well in cross-scenario jamming detection using real-world 5G measurements.

\subsubsection{\textbf{GANs with sequential models}}
Network traffic data, such as packet-level data, is inherently a time-series distribution. Using DL models that capture sequential properties (e.g., RNNs, LSTMs) significantly enhances the learning of time-series features. Lin \textit{et al.}\cite{lin2020using} use LSTM as generators within GANs to effectively capture correlations between points in the feature time series. Their approach includes min-max normalization to prevent mode collapse. The model also generates metadata as an auxiliary output, which serves as a conditional attribute for generating real feature values. In a later study, the same authors enhance this method by synthesizing packet and flow header data across multiple measurement epochs in~\cite{yin2022practical}. This improvement addresses the limitations of tabular-based GAN methods that fail to capture inter-epoch correlations. Their method generates both flow and packet-level data, with packet-level generation handled separately for each flow. Building on the approach proposed by~\cite{lin2020using}, the authors in~\cite{wang2024progen} adapt the model for network traffic generation in the context of malicious traffic detection. They train the model with both malicious and benign traffic to generate both types and analyze their distributions effectively.

Transformer models are also integrated with GAN models for improved performance. Zhang et al.~\cite{zhang2024transflowgan} use Relative GAN (Relativistic GAN) along with transformer modules. The Relative GANs enhance the discriminator by predicting the probability that a given sample is more realistic than other samples, rather than directly classifying a sample as real or fake. In this approach, the generator is not only responsible for producing realistic samples but also for maintaining the relative authenticity differences between samples. The relativistic discriminator improves training by reducing mode collapse and promoting diversity in generated outputs. Additionally, the internal structure of their model incorporates transformer modules to effectively capture temporal relationships, enhancing the generation of time-dependent data (e.g., network traffic).

The authors in~\cite{shahid2020generative} explored the integration of autoencoders and LSTMs with GANs to enhance generation quality, particularly for improving Network IDS (NIDS). In their framework, an autoencoder was first trained to learn a latent representation of real sequences of packet sizes.
This autoencoder consists of LSTM layers to capture temporal dependencies in sequential data, particularly the ordering of the packet sizes in a sequence. Subsequently, GAN was trained on this latent space to generate latent vectors, which could then be decoded into realistic sequences of packet sizes. In a similar approach, ~\cite{qu2024towards} propose a GAN model combined with a LSTM approach for IDS in fog computing network environments. After collecting network data, noise is removed, and features are filtered before generating  data using the GAN model. This processed data is then used to train LSTM for detecting anomalous traffic in fog computing networks.

\subsubsection{\textbf{Miscellaneous approaches}}
In this section, we review the literature that explores the use of other GAN variants and the integration of different model architectures.
In~\cite{anande2023generative}, the authors use Conditional Tabular GAN (CTGAN) and Copula GAN for feature generation and compare their performance to Vanilla GANs both quantitatively and qualitatively. They apply different data preprocessing techniques for each GAN type: Yeo-Johnson power transform~\cite{yeo2000new} for Vanilla GAN, the Reversible Data Transforms (RDT) library~\cite{montanez2018sdv} for CTGAN, and a Gaussian Copula transformer for Copula GAN. These techniques enhance training and generation performance. Their experiments show that up to 85\% of the generated (fake) data features can replace real data features without being detected, highlighting the effectiveness of their approach. Additionally, the authors in \cite{meddahi2021sip} employ an encoder-decoder architecture to convert packet bytes into images and use Deep Convolutional GANs (DCGANs) to generate images that enhance Session Initiation Protocol (SIP) traffic data.

The authors in~\cite{dowoo2019pcapgan} propose a model that combines multiple GAN architectures—Graph GAN, Image GAN, and Sequence GAN—to process different data types in parallel, generated by style-based encoders. Graph GAN processes graph data that encodes relationships between packet IP addresses. Image GAN takes image data created by converting time intervals into visual representations. Sequence GAN handles packet-layer information (e.g., Ethernet, IP) as numerical sequences. Additionally, they generate optional data, such as packet loads, using Sequence GAN and MMs. The synthetic data produced by these trained models is decoded to recreate \texttt{.pcap} files.

Yang et al.~\cite{yang2024research} generate network traffic using traffic obfuscation techniques. They use GMM for data normalization and autoencoders to reduce network data to a low-dimensional form. This processed data is then fed into a GAN model that integrates Swing Transformers. Swing Transformers are a specialized type of transformer architecture designed to capture feature variations at multiple scales within DL layers. The use of Swing transformers has enabled the model to capture feature variations across different scales within the DL layers while also integrating self-attention mechanisms.

Extending their research into practical usage, \cite{kim2024network} utilized CTGAN \cite{ctgan} for data generation to build a network traffic simulation platform. They began by decoding packet data to train the CTGAN model. The generated data was then encoded into synthetic packet information in the form of a \texttt{.pcap} stream. The quality of the generated data was assessed using a simulation setup by comparing it against real packet streams.

Khan et al.\cite{khan2024secure} proposed a data generation approach for Software-Defined Networking (SDN) settings in healthcare consumer IoT device management, specifically addressing class imbalance. The method utilizes a Boundary Equilibrium GAN (BEGAN), with a separate GAN model dedicated to each data class. The training process is halted based on a convergence metric to ensure optimal performance. Additionally, autoencoders are employed for dimensionality reduction in the data, enhancing the efficiency of the model.

In addition to the practical data generation, theoretical analysis of GAN-based data generation is also studied. For example,~\cite{lin2021privacy} explores the privacy implications of synthetic data generated by GANs, focusing on differential privacy and membership inference attack resilience. The authors prove that generated samples inherently have weak differential privacy properties, scaling with the ratio of generated to training samples. They also show that vanilla GANs, without special privacy mechanisms, are robust against black-box membership inference attacks. Their contributions lie in providing theoretical bounds on privacy risks for GAN-generated samples, highlighting areas where traditional GANs meet privacy requirements and where further techniques are needed.

\subsubsection{\textbf{Limitations in GAN based traffic generation}}

A significant challenge in GAN-based modeling is mode collapse, where the generator produces a limited variety of data, failing to capture the full range of variations in the training dataset~\cite{lin2020using}. This issue is particularly pronounced in network traffic generation, as traffic traces often encompass a diverse set of modes that are challenging for the generator to learn. However, this challenge can be effectively mitigated through approaches such as robust data normalization techniques and other advanced pre-processing methods~\cite{lin2020using}.
Another major challenge is training instability, which arises when the generator and discriminator fail to reach equilibrium during training~\cite{asperti2020balancing, ding2022take}. This instability can result in synthetic traffic traces with reduced fidelity.
Additionally, vanishing gradients is also a common issue in GANs when the discriminator becomes overly strong. In such cases, the generator receives weak gradient signals, hindering its ability to improve~\cite{ding2022take}. For network traffic generation, vanishing gradients can significantly affect the learning of long-term temporal dependencies in network traffic traces and lead to insufficient generalization of the data, as smaller gradients prevent the generator from learning comprehensive feature patterns.

\subsubsection{\textbf{Summary}}
GANs are among the most widely used generative AI models for network traffic generation. Table \ref{table:gan summary} provides a comprehensive summary of the work done with GANs. To optimize these models for traffic generation, various GAN variants have been proposed, particularly by modifying their internal neural network architectures. These adaptations include integrating autoencoders, RNNs, and incorporating conditional information. Such enhancements have successfully facilitated traffic generation for a wide range of network-related applications, including IDS and improving other machine learning tasks, primarily for both packet-level and flow-level data. While GANs inherently face challenges such as mode collapse and training instability, existing approaches have effectively addressed these issues through advanced mechanisms, enabling the generation of high-fidelity data.

\vspace{-4mm}
\subsection{Diffusion Models (DM)}

\begin{table*}[t!]
\scriptsize
    \centering
    \caption{Summary of DM-based work reporting application scenario, model architecture, and the data being used.}
    \label{table:dm summary}
    \begin{tabular}{|>{\centering\arraybackslash}m{1.6cm}|>{\centering\arraybackslash}m{0.6cm}|>{\centering\arraybackslash}m{4.3cm}|>{\centering\arraybackslash}m{3.3cm}|>{\centering\arraybackslash}m{1.4cm}|>{\centering\arraybackslash}m{3cm}|>{\centering\arraybackslash}m{0.8cm}|}
    \hline
    \multirow{2}{*}{\textbf{Model}} & \multirow{2}{*}{\textbf{Work}} & \multirow{2}{*}{\textbf{Application Scenario}} & \multirow{2}{*}{\textbf{Model}} & \multicolumn{2}{|c|}{\textbf{Data}} & \textbf{Artifacts} \\
    \cline{5-6}
    \textbf{Variant}&&&\textbf{details}&\textbf{OSI layer}&\textbf{Format/Type}&\textbf{available}\\
    \hline

    \multirow{1}{*}{DDPM}
    & ~\cite{sivaroopan2024netdiffus} & Anomaly detection, traffic fingerprinting & DDPM & Transport, Network & Packet level & \href{https://github.com/Nirhoshan/NetDiffus}{Yes} \\
    \cline{2-7}

    & ~\cite{rf_diffusion} & Generic RF signal generation & HDT & Physical & Raw signal data & \href{https://github.com/mobicom24/RF-Diffusion}{Yes} \\
    \cline{2-7}
    
    & ~\cite{wang2025privacy} & Privacy against MIA & RF-Diffusion & Physical & WI-FI signal & No \\ 
    \cline{2-7}
    
    & ~\cite{nute2025radio} & RF data generation & DDIM & Physical & RF signal & No \\ 
    \cline{2-7}
    
    & ~\cite{yin2025noise} & RFFI for IoT device authentication & DDPM & Physical & RF signal & No \\ 
    \cline{2-7}
    
    & ~\cite{yun2025smote} & Radar signal synthesis in intelligent transportation system & DM + SMOTE & Physical & Radar signal & No \\ 
    
    \hline
    
    \multirow{1}{*}{Stable DM}
    & ~\cite{jiang2024netdiffusion, jiang2023generative} & Generic network traffic generation & Stable DM + ControlNet~\cite{zhao2024uni} & Transport, Network & Packet level & \href{https://github.com/noise-lab/NetDiffusion_Generator}{Yes} \\
    \cline{2-7}
    
    & ~\cite{wang2025high} & High-resolution mmWave imaging & Stable DM & Physical & mmWave & No \\ 
    \hline

    \multirow{5}{*}{Controlled DM}
    & ~\cite{zhang2024netdiff} & Mobile app usage modeling and service interaction tracing & Hierarchical DM + Word2Vec + Encoder-Decoder + Two-layer transformer & Transport, Application & Network flow traces, App usage sequences & No \\
    \cline{2-7}
    & ~\cite{chai2024diffusion} & Network optimization & DM + Attention-based classifier-free guidance + Contrastive learning & Application & Aggregated mobile traffic data across geographic areas & \href{https://github.com/tsinghua-fib-lab/opendiff}{Yes} \\
    \cline{2-7}
    & ~\cite{chai2024knowledge} & Mobile traffic generation, Network optimization & Conditional DM + KG + Frequency attention & Application & Mobile traffic time-series, Urban data & No \\
    \cline{2-7}
    &\cite{qi2024regional}& 5G network traffic generation & CANDLE + GCN & Application & Traffic time-series, Geospatial grid-based traffic patterns &   No \\
    \cline{2-7}
    & ~\cite{jasnidiffupac} & Adversarial packet generation, Evasion of NIDS & DM + BERT & Transport, Network & Flow level & No \\
     \cline{2-7}
    & ~\cite{11004012}& Secured sensing system & Conditional DM & Physical & CSI data &   No \\
     \cline{2-7}
    & ~\cite{fu2025diffusion}& Denoising AE signals in concrete & Conditional DM + CNN + Attention & Physical & AE signal &   No \\
    \cline{2-7}
    & ~\cite{cheng2024wivid}& Depth estimation & Conditional DM  & Physical & CSI data &   \href{https://github.com/Guoxuan-Chi/WiViD}{Yes} \\
    \cline{2-7}
    & ~\cite{11059502}& ISAC & Conditional DM  & Physical & CSI data &   \href{https://github.com/Guoxuan-Chi/WiViD}{Yes} \\
    \hline

    \multirow{1}{*}{Lightweight DM}
    & ~\cite{li2024lightweight} & Synthesis of malicious traffic for edge devices & DM with depthwise separable convolutions & Network & Packet level & No \\
    \hline

    \end{tabular}
\end{table*}

DMs were first introduced to the DL community in 2015 by~\cite{sohl2015deep}, where they were initially conceptualized as probabilistic generative models. However, it was not until 2020, with the groundbreaking work in~\cite{NEURIPS2020_4c5bcfec}, that DMs gained significant traction. The development
of these models
marked a turning point in generative modeling, with DMs quickly emerging as a compelling alternative to long-dominant GANs.

\begin{figure}[h!]
\centering
\includegraphics[width=\linewidth]{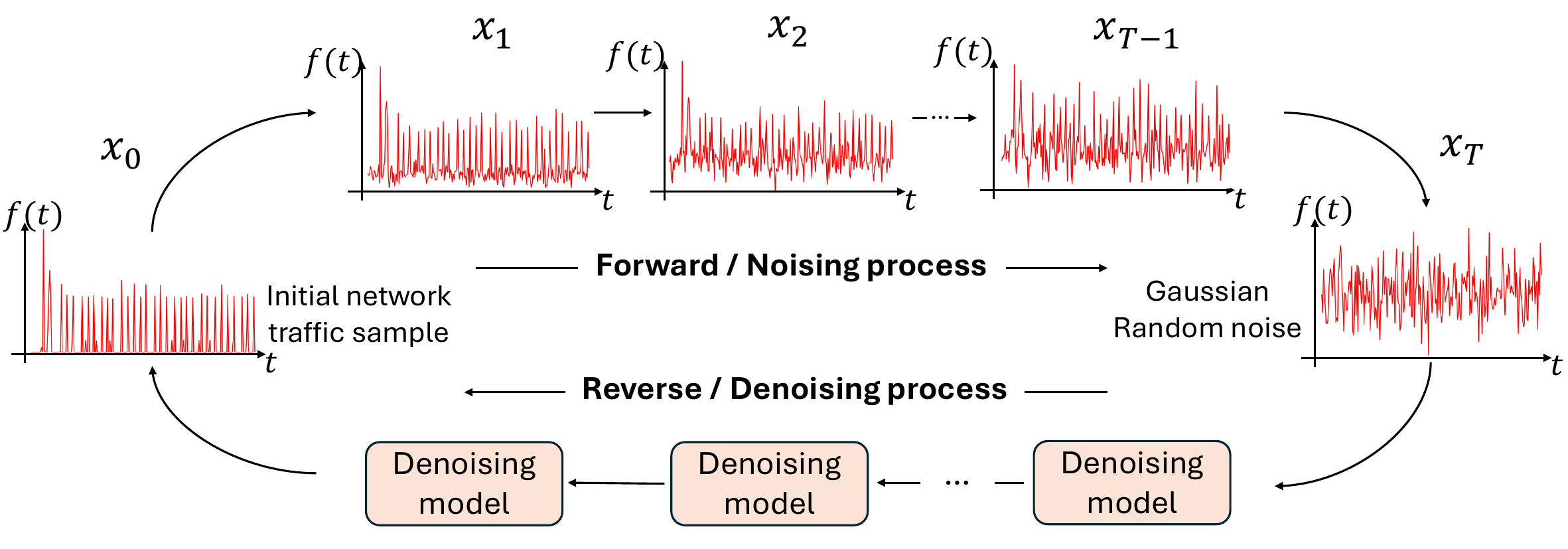}
\caption{Basic functionality of DM}
\label{fig:DMs}
\end{figure}

DMs operate based on a probabilistic framework that transforms random noise into meaningful data samples. As shown in Fig.~\ref{fig:DMs}, this is achieved through a two-step process. 
First, the forward process gradually adds Gaussian noise into the input sample ($x_0$) until $T$ time steps, creating a series of noisy data representations from ($x_1$ to $x_T$). The model is then trained to reverse this process, step by step, by learning the conditional probabilities that describe how to transform a noisy version of the data back into a less noisy version. The DNN model, a.k.a. denoising model, which conducts this transformation, learns the underlying noise probability distribution of the data. Compared to VAE and GAN, the iterative denoising process in DM provides more stability during the model training while offering a diverse representation of the synthetic data, mitigating issues such as mode collapse.
Different extensions of DM have been proposed so far for network traffic generation as we have summarized in Table~\ref{table:dm summary} and explained next.

\subsubsection{\textbf{Denoising  Diffusion Probabilistic Model (DDPM)}}

DDPMs optimize the denoising process by explicitly modeling the noise distribution, allowing for sharper and more diverse samples. DMs were first adapted for network traffic generation in~\cite{sivaroopan2024netdiffus}, where the authors employed DDPM to model time series data representing packet-level features, specifically downloaded byte streams. In addition to time series data, they also utilized GASF images, which were derived from the same time series data. A comparison was drawn with the SOTA GANs that had previously been used for network traffic generation, focusing on both 1D features and 2D images. Their results demonstrated a significant improvement in the accuracy of network traffic image generation using DMs, outperforming previous GAN-based methods, particularly in cases that leveraged packet-level features in 2D image format.

Generating high-fidelity RF signals presents unique challenges due to the need to capture their full structure across the time, frequency, and complex value domains. In ~\cite{rf_diffusion}, the authors highlight that RF signals, as time-series data, capture dynamic details that standard diffusion models struggle to synthesize. Moreover, critical RF features embedded in the frequency domain, such as Doppler shifts, are often overlooked due to the focus of DMs on spatial and amplitude information. Furthermore, the complex-valued nature of RF signals, including phase data, is neglected in existing models. Addressing these complexities, the authors introduce RF-Diffusion to synthesize RF signals utilizing a novel Time-Frequency Diffusion (TFD) theory and a Hierarchical Diffusion Transformer (HDT). In TFD theory, they present RF-Diffusion model which alternates between adding noise in the time domain and blurring the frequency domain during the forward process and reverses these effects by denoising and deblurring. This restoration emphasizes the accuracy of the amplitude in the time domain and the continuity of the frequency domain. The HDT incorporates a hierarchical architecture to decouple spatio-temporal dimensions, attention-based diffusion blocks with enhanced transformers for feature extraction, and a complex-valued design to encode both amplitude and phase information. The authors synthesize Wi-Fi and FMCW signals using RF-Diffusion which outperforms typical DDPM, DCGAN, and CVAE-based methods.

Building on top of the RF-diffusion model, authors of ~\cite{wang2025privacy} added differential privacy (DP) to the base model for privacy-preserving WI-FI data generation.They had proposed a hybrid training method for the DM applied to wireless data as a defense against Membership Inference Attacks (MIA). The approach involves initially training the model without privacy constraints. After a specified number of training rounds, DP
is incorporated for fine-tuning. During this second phase, a cooptimization  process is conducted in parallel to counteract the effects of the added noise.   

In ~\cite{nute2025radio}, the authors propose a DM tailored for RF data generation using Denoising Diffusion Implicit Models (DDIM) with noise schedules optimized for In-phase/Quadrature (I/Q) signal types. Unlike DDPM, this approach reduces computational overhead while maintaining high-quality output. In ~\cite{yin2025noise}, the authors leverage DM to enhance RF Fingerprint Identification (RFFI) for IoT device authentication under low SNR conditions. By training a robust noise predictor and designing a noise removal algorithm, the approach effectively denoises received signals, restoring subtle hardware features crucial for device identification. This was demonstrated on Wi-Fi signal. In ~\cite{yun2025smote}, the authors propose SMOTE-Diffusion, a DM combined with the SMOTE to improve stationary object classification using Impulse Radio UltraWideBand (IR-UWB) radar. While FMCW radar excels at detecting moving objects, IR-UWB offers high temporal resolution suitable for stationary object recognition but suffers from limited time-domain data. SMOTE-Diffusion addresses this by generating diverse, realistic synthetic radar signals, reducing data collection needs and enhancing model performance.

\subsubsection{\textbf{Stable DM}}

Stable DM generates controlled high-quality images from textual descriptions, enabling both text-to-image and image-to-image transformations. Further, Jiang \textit{et al.} \cite{jiang2024netdiffusion, jiang2023generative} applied stable DMs to address the problem of network data scarcity. The authors converted network packet traces into images through the use of nPrint, a method that standardizes bit-level information from packet headers into a consistent format. These images were then utilized with a stable diffusion model to synthesize network traffic, offering a controlled synthesis process. A key innovation in this work was the integration of ControlNet\cite{zhao2024uni}, which provided an additional layer of control, ensuring that essential network characteristics, such as the number of bits per header, were preserved in the synthesized images. To further enhance the realism of the generated data, a post-processing mechanism was introduced to maintain adherence to protocol-specific constraints and characteristics.

In ~\cite{wang2025high}, the authors propose a diffusion-based neural network for high-resolution mmWave imaging, which transforms mmWave signals into high-quality images by leveraging the intrinsic features of target scenes. 
This approach uses conditional generation through stable diffusion to enhance image quality without relying on sparsity assumptions. To support this, they design an optimized mmWave metasurface and codebook that improve signal penetration and phase control, enabling effective imaging with compact, off-the-shelf hardware and no mechanical movement.

\subsubsection{\textbf{Controlled DM}}

Expanding the scope of diffusion models in network traffic generation and cybersecurity, recent research has focused on improving controllability, reducing data dependency, and capturing complex spatial-temporal patterns – key areas where traditional GAN-based approaches often fall short. Zhang \textit{et al.} ~\cite{zhang2024netdiff} presented a hierarchical, service-guided, controlled diffusion model for network flow trace generation. This approach was designed to explicitly model users' dynamic network usage patterns, focusing on the interactions between users and network services, such as mobile application usage traces. Their hierarchical generation structure consisted of multiple layers, the first of which was responsible for modelling app usage behaviour. This app usage data was then used to guide the generation of network flow traces, ensuring that the generated flows reflected realistic service usage patterns. The model captured co-usage relationships between different services, as well as sequential usage patterns, by leveraging a pre-trained embedding model and an encoder-decoder architecture. Furthermore, it incorporated a two-layer transformer network to model the temporal and feature correlations inherent in multidimensional network flow data. The generation process followed a two-step structure: first, a diffusion model generated the app usage patterns, and subsequently, a second diffusion model was trained on network flow traces, using the app usage patterns to guide the generation of realistic network traffic data.

In~\cite{chai2024diffusion}, the authors introduce OpenDiff (i.e., for satellite image data), Point of Interest (POI) distributions, and population density to generate mobile traffic data for network planning and optimisation. By adopting a multi-positive contrastive learning algorithm and an attention-based classifier-free guidance module, OpenDiff mitigates the reliance on non-public data and enhances the controllability of traffic generation. Data pre-processing steps included feature extraction from satellite images and POI, population distribution analysis, and multi-positive sample construction for contrastive learning. 

The paper, CANDLE~\cite{qi2024regional}, uses a conditional diffusion model to simulate 5G network traffic in regions with limited historical data, conditioning on existing 4G traffic patterns. This model incorporates graph convolutional networks (GCN) and a cross-attention mechanism to address data scarcity and provide operators with insights for base station deployment. In addition, in~\cite{chai2024knowledge}, KG-Diff is proposed to incorporate urban contextual features from an Urban Knowledge Graph (UKG) to guide mobile traffic generation. By employing a frequency attention mechanism that decomposes network traffic into multiple periodic components, KG-Diff effectively models complex temporal patterns and the underlying correlations between urban environments and network traffic, offering a more nuanced approach to mobile traffic generation. The Fourier transform was used for frequency decomposition.

Extending the application of DMs to cybersecurity, DiffuPac~\cite{jasnidiffupac} innovatively combines a Bidirectional Encoder Representations of Transformers (BERT) model with a diffusion framework to generate adversarial network packets capable of evading detection by NIDS. DiffuPac introduces a targeted noising mechanism and classifier-free guidance, ensuring that only malicious packet sequences are noised while preserving the integrity of normal traffic. Additionally, the model employs a unique packet concatenation strategy to seamlessly integrate malicious packets into benign traffic flows, leveraging contextual embeddings from BERT for enhanced mimicry. DiffuPac's fine-tuning phase optimises a variational lower bound that prioritises the reconstruction of malicious packets as indistinguishable from normal traffic patterns. Evaluations demonstrate that DiffuPac not only outperforms traditional approaches like GANs and LSTMs but also operates under the realistic constraint of limited attacker knowledge, showcasing its practical efficacy in real-world scenarios.

In ~\cite{11004012}, a DM-based secure sensing system is proposed to enhance the security of integrated sensing and communications by preventing unauthorized sensing using CSI. The system uses discrete and continuous conditional diffusion models to selectively activate network elements and generate safeguarding signals that mask user activity. This ensures that only authorized devices can recover accurate CSI for sensing. AEDM~\cite{fu2025diffusion}, a DM-based DL approach for denoising acoustic emission (AE) signals in concrete, aimed at improving structural health monitoring. The model combines a CNN with attention mechanisms and multi-scale convolutions, and conditions the diffusion process on noise observations to effectively separate noise from relevant signals. This enables accurate damage detection despite challenges like material anisotropy and external interference. WiViD~\cite{cheng2024wivid}, a diffusion-based depth estimation system that combines commercial Wi-Fi signals with vision to overcome the limitations of Light Detection and Ranging (LiDAR), mmWave, and monocular vision methods. By introducing a Multimodal Conditional Diffusion (MMCD) mechanism, WiViD iteratively refines depth predictions for improved accuracy. The system features two key encoders—the Complex-Valued CSI Encoder (CCE) and the Residual Image Encoder (RIE)—to effectively capture and fuse spatio-temporal features from Wi-Fi CSI and RGB images. This fusion enables robust and high-precision depth estimation suitable for real-world applications. In ~\cite{11059502}, the authors propose a DM-based data augmentation system to address the challenge of limited CSI data in Integrated Sensing and Communication (ISAC) systems. The approach trains a conditional DM on real-world samples to generate additional CSI data, increasing sample quantity. A second DM is then applied to denoise the generated data to improve quality. This method supports more effective AI-driven CSI analysis for physical space and human activity monitoring in 6G networks.

\subsubsection{\textbf{Light-weight DMs}}

Another angle of improvement for DMs were addressed in making the model lightweight such that it can be used for high-quality data generation at edge devices. \cite{li2024lightweight} proposes a DM specifically designed for resource-constrained edge devices. By integrating depthwise separable convolutions into the diffusion process, this model achieves significant reductions in computational overhead while maintaining the diversity and quality of generated malicious traffic. The evaluations demonstrate the ability of the model to produce high-quality synthetic data, improving the training of IDS classifiers even on hardware-limited edge nodes. This lightweight design underscores the potential of diffusion models to address computational challenges, enabling real-time cybersecurity applications at the edge.

\subsubsection{\textbf{Limitations}}

DMs face challenges in traffic generation, including limited controllability under specific conditions and difficulties in handling multimodal data or generating realistic spatial and temporal patterns. Traditional DMs often lack mechanisms to incorporate diverse contextual features effectively, reducing fidelity and adaptability. To address these issues, frameworks like OpenDiff~\cite{chai2024diffusion} employ classifier-free guidance and attention-based fusion to enhance controllability and integrate conditional features such as human activity data. CANDLE~\cite{qi2024regional} uses cross-attention mechanisms and graph-based architectures to align 4G traffic data with 5G demand, improving adaptability in underdeveloped regions.

Additionally, DMs have high computational demands, making them unsuitable for resource-constrained edge devices~\cite{li2024lightweight}. Models like KG-Diff~\cite{chai2024knowledge} address this by incorporating UKG and frequency attention mechanisms to capture urban context and multi-scale periodicities, enhancing fidelity and controllability. DiffuPac~\cite{jasnidiffupac} integrates BERT and targeted noising strategies to generate adversarial packets that blend seamlessly into normal traffic while maintaining functionality. Lightweight designs, such as depthwise separable convolutions, reduce computational overhead, enabling deployment on edge devices. These advances improve the practicality and effectiveness of DMs in traffic synthesis and cybersecurity.

\subsubsection{\textbf{Summary}}
The emergence of DMs has revolutionized the field of network traffic analysis, providing a powerful alternative to traditional generative models like GANs. 
 Table \ref{table:dm summary} provides a comprehensive summary of the work done with DMs. Innovative adaptations of DMs, such as DDPM, stable diffusion models, and controlled diffusion models, have demonstrated remarkable capabilities to generate realistic high-quality network traffic data. By effectively modeling the underlying distributions of packet-level features and user interactions with network services, DMs have consistently outperformed previous SOTA GANs in terms of accuracy and realism. The ability of DMs to capture complex temporal and feature correlations while mitigating issues such as mode collapse has solidified their position as the leading generative framework for network traffic synthesis.  Further, lightweight diffusion models make the way for high-quality data synthesis in a resource-constrained environment. As research in this area continues to advance, DMs are increasingly pivotal for network security, traffic data analysis, and addressing data scarcity in network environments.

\vspace{-4mm}
\subsection{Transformers}

\begin{table*}[t!]
\scriptsize
    \centering
    \caption{Summary of Transformer based work reporting application scenario, model architecture and the data being used.}
    \label{table:transformers summary}
    \begin{tabular}{|P{0.7cm}|P{5.5cm}|P{3cm}|P{2.5cm}|P{2.75cm}|P{1cm}|}
    \hline
    \multirow{2}{*}{\textbf{Work}}& \multirow{2}{*}{\textbf{Application scenario}}& \multirow{2}{*}{\textbf{Model(s) used}}& 
    \multicolumn{2}{|c|}{\textbf{Data}}&\textbf{Artifacts}\\
    \cline{4-5}
    &&&OSI layer&Format/Type&\textbf{available}\\
    \hline
       
        \cite{bikmukhamedov2020generative}& Generic network traffic generation, Attack detection & GPT-2  & Transport, Network &Packet and Flow level& No\\
         \hline
         \cite{meng2023netgpt}& Generic network traffic generation  & GPT-2   & Transport, Network & Packet and Flow level& No\\
         \hline
         \cite{qu2024trafficgpt}& Network optimization and traffic flow classification & GPT + Linear attention & Network & Flow level &  No \\
        \hline
        \cite{kholgh2023pac}& Generic network traffic generation, IDS & GPT-3 & Transport, Network & Packet and Flow level & No\\
        \hline
        \cite{kong2024high}&Cellular network traffic synthesis& CPT-GPT& Application& Event-level sequences & No\\
        \hline
        \cite{wang2025generative}& 3-D Human pose estimation & LDT & Physical& RFID signals & No\\
        \hline
    \end{tabular}
\end{table*}

Transformers were first introduced in 2017 in~\cite{vaswani2017attention} to address sequence transduction tasks, such as neural machine translation. These tasks involve converting an input sequence into an output sequence. The transformer model is a type of neural network that learns the context of sequential data and generates new data based on its understanding of the relationships between elements within the sequence.
Transformers uses attention mechanism to understand context weighing the importance of different elements in a sequence. Table \ref{table:transformers summary} summarizes recent work with Transformers

\begin{figure}[h!]
\centering
\includegraphics[width=0.9\linewidth]{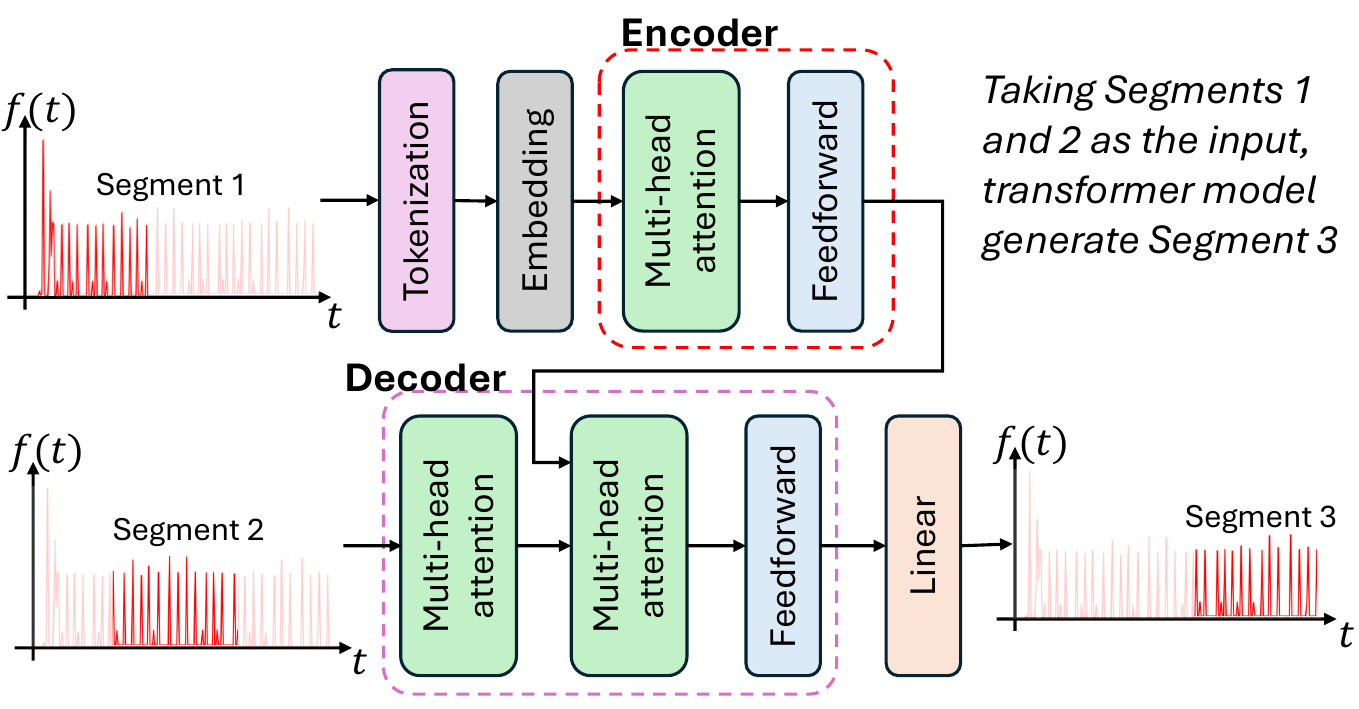}
\caption{Basic functionality of Transformers}
\label{fig:Transformers}
\end{figure}

As shown in Fig.~\ref{fig:Transformers}, tokenization, embedding, encoder, and decoder (i.e., with multi-head attention and feedforward layers) are the components of a generic transformer module used for network traffic generation~\cite{vaswani2017attention}. Taking network traffic generation as an example, this process starts by tokenizing the network traffic trace into smaller units, which are converted into vector representations through an embedding layer. The encoder processes these embeddings using multi-headed attention and feedforward layers, capturing both global and local dependencies in the traffic sequence. The decoder, which takes another segment of the network traffic trace as input, uses multi-headed attention to integrate context from the encoder and applies feedforward layers to generate the subsequent traffic sequence. This setup effectively models and reproduces the patterns in network traffic data.

\subsubsection{\textbf{Generative Pretrained Transformers (GPT)}}
Generative Pretrained Transformers (GPT) are a specific class of transformer models initially developed for natural language processing (NLP). Their utility extended beyond NLP, finding applications in various fields, including network traffic generation. After GPT-2 was released, it was leveraged in 2020 by ~\cite{bikmukhamedov2020generative} for synthetic network traffic generation and classification. The model treated packet size and inter-packet time intervals as two-dimensional features, mapped to a one-dimensional source space using K-Means clustering. The cluster embeddings resembled tokens used in transformers for training. By feeding the traffic class as the first token into the pre-trained model, it could generate a flow corresponding to that class. 

Further advancements were made in ~\cite{meng2023netgpt}, where the authors developed NetGPT, a GPT model pre-trained on extensive network traffic data irrespective of the downstream task that needs to be performed. During fine-tuning, task-specific labels (e.g., software identification, source port generation) were added. To enhance the dataset for fine-tuning, they shuffled packet headers, as the order of headers was irrelevant for certain tasks. However, the token limit posed challenges for flow length during training and generation. To mitigate this, ~\cite{qu2024trafficgpt} introduced TrafficGPT, which employed a linear attention mechanism to extend the token scope from 512 to 12,032 tokens. They also developed a reversible token representation method to enable bidirectional mapping between pcap files and tokenized representations. Metadata like MAC and IP addresses were removed to avoid overfitting.

With the introduction of GPT-3 models, authors of ~\cite{kholgh2023pac} leveraged GPT-3 Davinci model and GPT-3 Babbage model to synthesize network traffic for various tasks. The pipeline of this process would be triggered by a user requesting network traffic either using a specific scenario (e.g., normal traffic data, a Ping-of-Death attack, etc.) or by specifying a set of network protocols (e.g., Internet Control Message Protocol [ICMP], Domain Name System [DNS], etc.) Then, a Flow Generator, which is a Python script, produces network flow in text format, i.e. it generates the sequence and textual summary of packets that make up the network traffic, although not creating the actual packets themselves. The flow summary is then passed to the Packet Generator, which is a transformer-based model capable of creating packets using the network flow. The packet generator then creates the specified packets and writes them in \texttt{.pcap} format to a file.GPT-3 Davinci generated training samples for accuracy, while GPT-3 Babbage powered the packet generator for cost-efficient and faster inference.

Similarly, ~\cite{kong2024high} presents CPT-GPT, a high-fidelity traffic generator tailored for cellular network control-plane traffic. CPT-GPT effectively addresses challenges such as stateful semantic correctness, multimodal feature relationships, and long-term data drifts by leveraging a Transformer-based architecture. Through its innovative tokenization scheme and probabilistic output mechanism, CPT-GPT generates realistic traffic while avoiding issues like state violations and mode collapse. Unlike traditional Semi-Markov models (SMMs) that require extensive manual updates to align with evolving the 3rd Generation Partnership Project (3GPP) standards, CPT-GPT  achieves better fidelity. 

The work in ~\cite{wang2025generative} relies on pretrained GPT models and instead introduces a novel framework using latent diffusion transformers (LDT) to generate high-quality, diverse RFID sensing data. Addressing the challenges of collecting paired RFID and 3D human pose data, the synthetic data is used to train a transformer-based kinematics predictor for temporally smooth 3D pose estimation. The framework includes a cross-attention conditioning stage during training and a two-stage velocity alignment design at inference to accurately infer missing joints. This enables the completion of full 25-joint skeletal poses from only 12-joint inputs.

\subsubsection{\textbf{Limitations}}

Transformer (e.g., GPT) models face limitations in network traffic tasks due to the heterogeneous syntax of headers and payloads, especially with encrypted traffic, which complicates vocabulary construction and semantic retention. Their quadratic complexity with input length poses challenges for long packet flows, while unidirectional models struggle with bidirectional tasks like classification and attack detection. The limitations in token length and the inability to accurately reconstruct realistic traffic patterns further hinder their effectiveness~\cite{qu2024trafficgpt}. To address these, NetGPT~\cite{meng2023netgpt} employs general encoding with hexadecimal tokens, processes only the first three packets of heavy flows to reduce overhead, and improves adaptability via header field shuffling and packet segmentation. TrafficGPT~\cite{qu2024trafficgpt} enhances the handling of long sequences with linear attention mechanisms (supporting up to 12,032 tokens) and reversible token representations for precise traffic reconstruction. Together, these strategies improve the utility of transformer models in various traffic understanding and generation tasks~\cite{meng2023netgpt, qu2024trafficgpt}.

\subsubsection{\textbf{Summary}}

The integration of transformer models, particularly GPT, into network traffic generation marks a major advance in AI-driven data synthesis. Evolving from natural language processing to specialized frameworks like NetGPT and TrafficGPT, these models demonstrate strong versatility in handling complex network traffic data. Advances in tokenization and attention mechanisms address challenges such as flow length limits and class imbalance, improving the quality of synthetic traffic. As they mature, transformer models are expected to transform network analysis, strengthen security, and provide deeper insights through more accurate traffic generation.

\section{Comparative Analysis of Different Methods}
\label{sec:statsvsdl}

\begin{figure}
    \centering
    \includegraphics[width=\linewidth]{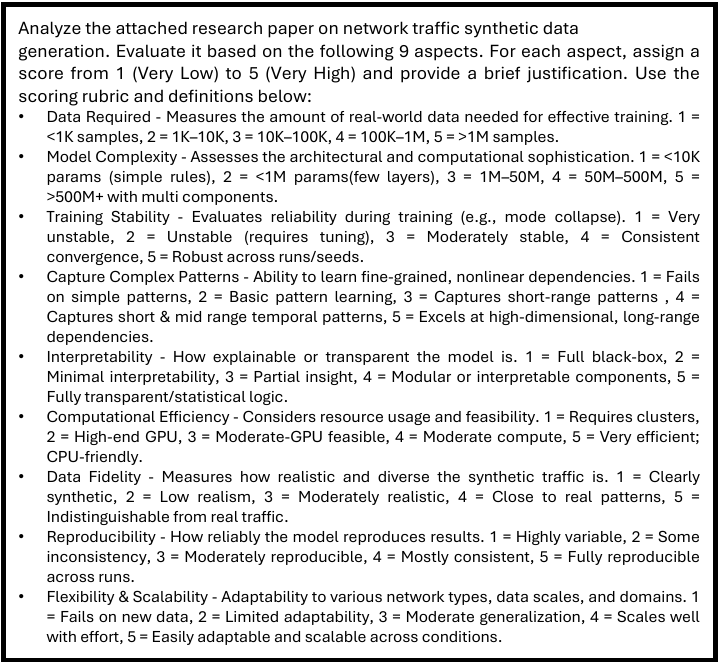}
    \caption{Prompt for the AI Tool. The prompt includes the rubric for the evaluation.}
    \label{fig:ai_prompt}
    
\end{figure}

To systematically assess generative approaches for network traffic synthesis, we evaluated each method across nine key aspects using a standardized 1–5 scoring rubric (see Fig.~\ref{fig:ai_prompt}). A score of 1 indicates the weakest performance or highest cost/limitation, while 5 reflects ideal performance or efficiency. Where possible, the criteria are grounded in measurable metrics such as number of parameters, required data size, training stability, and computational demands. 

\begin{figure}
    \centering
    \includegraphics[width=0.95\linewidth]{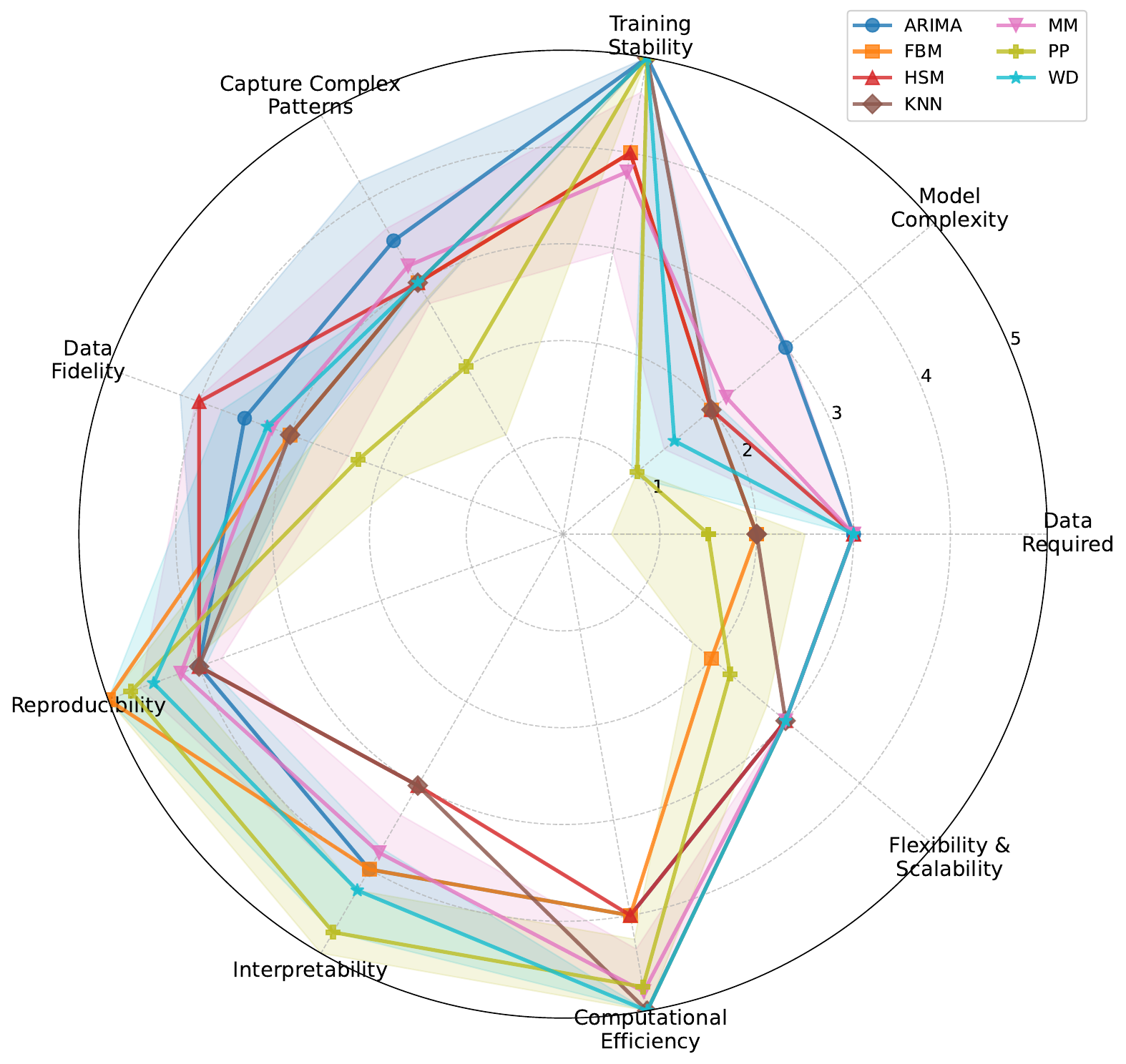}
    \vspace{-2mm}
    \caption{Comparison of synthetic traffic generation statistical models. Markers: mean scores, shaded region: standard deviation.}
    \label{fig:radar_comparison_statistical}
    
\end{figure}

\begin{figure}
    \centering
    \includegraphics[width=0.95\linewidth]{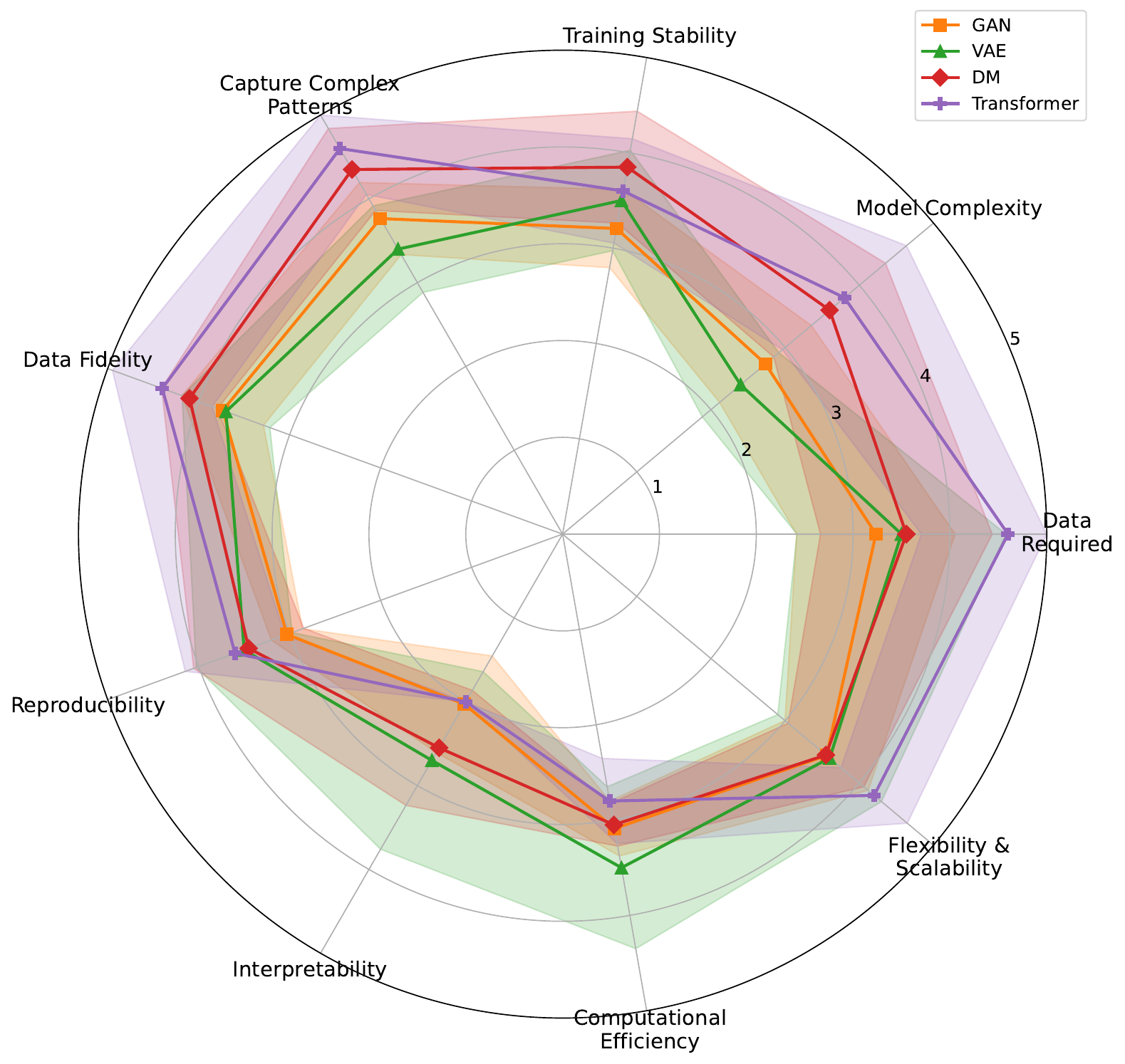}
    \caption{Comparison of synthetic traffic generation DL models. Markers: mean scores, shaded region: standard deviation.}
    \label{fig:radar_comparison_dl}
    
\end{figure}

We, the authors, selected 86 papers spanning all approaches discussed in this survey and scored them for each aspect using the provided rubrics, drawing on both our expertise and the findings from our literature review. The mean and standard deviation of these scores were then calculated and visualized in Fig. \ref{fig:radar_comparison_statistical} and Fig. \ref{fig:radar_comparison_dl}. To enable readers to quickly obtain scores and analyses for any additional paper they encounter, we provide a prompt (shown below) that can be run on an AI model using the extracted text from the target paper. From our labeled set, we selected random 20 papers as a training set and refined the prompt until it achieved 100\% agreement with our manual scores. We then applied the automated script to the remaining 66 papers to verify alignment, achieving 100\% accuracy with the final prompt provided in Fig.~\ref{fig:ai_prompt}.

For privacy reasons—avoiding the need to upload research papers to cloud-based models—we used GPT-OSS~\cite{openaiIntroducingGptoss} in a local setup. Readers can similarly run the provided prompt on a local model to obtain scores for any synthetic data generation paper. To demonstrate how the rubric applies with the provided prompt, Table~\ref{table:netdiffus_evaluation} presents a model-specific evaluation of \textit{NetDiffus}~\cite{sivaroopan2024netdiffus} using the proposed 9-aspect scoring scheme.

\begin{table}[!ht]
\centering
\scriptsize
\caption{Evaluation of ~\cite{sivaroopan2024netdiffus} across the 9 scoring aspects }
\label{table:netdiffus_evaluation}
\begin{tabular}{|@{}p{0.21\columnwidth} |c |p{0.60\columnwidth}|@{}|}
\hline
\textbf{Aspect} & \textbf{Score} & \textbf{Justification} \\
\hline
Data Required & 2 & Uses approximately 2K samples; strong performance observed with as few as 1K–10K traces. \\
\hline
Model Complexity & 3 & Implements a 5-layer U-Net with 1K diffusion steps; moderately complex, under 50M parameters. \\
\hline
Training Stability & 4 & Diffusion training is stable and consistent. \\
\hline
Capture Complex Patterns & 4 & GASF transformation and iterative denoising allow learning of temporal and nonlinear dependencies. \\
\hline
Interpretability & 2 & GASF inputs are interpretable, but the diffusion model itself remains largely a black box. \\
\hline
Computational Efficiency & 3 & Requires GPU acceleration; feasible on a single GPU but not CPU-friendly. \\
\hline
Data Fidelity & 4 & Synthetic traffic closely matches real traces; achieves significantly lower FID. \\
\hline
Reproducibility & 4 & DMs yield consistent outputs; high reproducibility across setups is likely. \\
\hline
Flexibility \& Scalability & 4 & Evaluated on diverse traffic types; generalizable with moderate effort. \\
\hline
\end{tabular}
\end{table}


\subsubsection{Comparison of Statistical Generative Approaches}

To provide a clear understanding of the relative capabilities of various generative approaches for network traffic synthesis, we present a radar plot (Fig.~\ref{fig:radar_comparison_statistical}) that visualizes key evaluation aspects.
As shown in the figure, statistical approaches consistently exhibit strong training stability, interpretability, and reproducibility, which stem from their explicit mathematical formulations and well-understood modeling assumptions. These properties make them well suited for controlled evaluations and analytical studies where transparency and robustness are critical. At the same time, the radar plot reveals clear limitations in capturing complex traffic patterns and achieving high data fidelity, particularly under highly dynamic or non-stationary conditions. Simpler processes such as PP and FBM demonstrate low flexibility \& scalability, while more advanced models, including ARIMA and HSM, provide moderate improvements by incorporating additional structure at the cost of increased model complexity. Across the category, computational efficiency remains a distinguishing advantage, enabling practical deployment in resource-constrained environments. Overall, Fig.~\ref{fig:radar_comparison_statistical} highlights the strengths of statistical generators in stability and efficiency, alongside their inherent expressiveness limitations.

\subsubsection{Comparison of DL Generative Approaches}

Fig.~\ref{fig:radar_comparison_dl} compares representative DL–based traffic generation models using the same evaluation aspects, enabling a direct interpretation of their modeling behavior. Unlike statistical approaches, DL models demonstrate strong performance in capturing complex traffic patterns and achieving high data fidelity, reflecting their ability to learn rich, non-linear representations directly from data. Transformer-based models, in particular, exhibit high flexibility \& scalability, driven by attention mechanisms that effectively model long-range dependencies. These gains, however, are accompanied by increased data requirements and model complexity, as evident from the corresponding axes in the radar plot. GANs show comparatively lower training stability, consistent with the challenges associated with mode collapsing. VAEs provide a more stable and computationally efficient alternative but exhibit reduced expressiveness relative to other DL architectures. DMs and Transformers provider higher data fidelity by capturing complex patterns. Across all DL models, interpretability remains limited, underscoring a recurring trade-off between modeling power and transparency. Collectively, Fig.~\ref{fig:radar_comparison_dl} emphasizes the suitability of DL-based generators for data-rich scenarios where realism and scalability are prioritized.

\subsubsection{Statistical vs. DL-Based Synthetic Network Traffic Generation}

Building on the family-specific observations in Figs.~\ref{fig:radar_comparison_statistical} and \ref{fig:radar_comparison_dl}, Table~\ref{table:stat_vs_dl_conceptual} provides a consolidated, high-level comparison between statistical and DL–based traffic generation paradigms. Rather than reporting quantitative scores, the table abstracts the dominant trends observed across both radar plots to highlight paradigm-level trade-offs. Statistical methods consistently emphasize interpretability, data efficiency, training robustness, and computational efficiency due to their explicit formulations and relatively low model complexity. In contrast, DL–based approaches prioritize expressiveness, flexibility \& scalability, and data fidelity by learning implicit representations from data, at the expense of transparency and efficiency. This synthesis clarifies that the choice between statistical and DL-based generators is fundamentally application-driven, depending on whether stability and interpretability or expressive power and realism are the primary design objectives.

\begin{table}[!ht]
\centering
\scriptsize
\caption{Conceptual comparison of statistical and DL-based synthetic traffic generation models. 
$\bullet$: strong advantage, $\circ\!\bullet$: mixed or context-dependent, $\circ$: clear limitation.}
\label{table:stat_vs_dl_conceptual}
\begin{tabular}{|p{0.32\columnwidth}|c|c|}
\hline
\textbf{Aspect} 
& \textbf{Statistical Methods} 
& \textbf{Deep Learning Methods} \\
\hline
Data Required & $\bullet$ & $\circ$ \\
\hline
Model Complexity & $\bullet$ & $\circ$ \\
\hline
Training Stability & $\bullet$ & $\circ\!\bullet$ \\
\hline
Capture Complex Patterns & $\circ$ & $\bullet$ \\
\hline
Interpretability & $\bullet$ & $\circ$ \\
\hline
Computational Efficiency & $\bullet$ & $\circ$ \\
\hline
Data Fidelity & $\circ\!\bullet$ & $\bullet$ \\
\hline
Reproducibility & $\bullet$ & $\circ\!\bullet$ \\
\hline
Flexibility \& Scalability & $\circ$ & $\bullet$ \\
\hline
\end{tabular}
\end{table}

\subsubsection{Constraint-Driven Model Selection Matrix}

Rather than selecting a generative model based solely on isolated performance metrics, practitioners should align model choice with the dominant constraints of the target application. Drawing on the comparative trends observed in Figs.~\ref{fig:radar_comparison_statistical} and~\ref{fig:radar_comparison_dl}, we summarize a constraint-driven matrix in Table~\ref{table:constraint_hierarchy} that maps common practical requirements to appropriate classes of synthetic network traffic generators.

\begin{table}[!ht]
\centering
\scriptsize
\caption{Constraint-driven matrix for selecting synthetic network traffic generators.}
\label{table:constraint_hierarchy}
\begin{tabular}{|p{0.28\columnwidth}|p{0.62\columnwidth}|}
\hline
\textbf{Primary Constraint} & \textbf{Recommended Model Classes and Rationale} \\
\hline
Interpretability, data efficiency, analytical transparency
&
Statistical models are preferred due to explicit formulations, low data requirements, stable training, and strong reproducibility. Simple processes (e.g., PP, FBM) suit baseline or stationary traffic, while structured models (e.g., MM, HSM, WD) better capture burstiness and long-range dependence without sacrificing interpretability.
\\
\hline
Moderate realism under limited data or compute
&
VAEs provide a balanced trade-off between realism and efficiency, offering stable training, moderate data fidelity, and feasibility on limited computational resources. Suitable when statistical models lack expressiveness but heavier DL models are impractical.
\\
\hline
High-fidelity and fine-grained traffic realism
&
DMs are preferred for their strong ability to capture complex temporal and nonlinear dependencies with stable optimization. GANs may be used when faster sampling is required, but their training instability and mode collapse risk must be carefully managed.
\\
\hline
Large-scale, heterogeneous, or evolving traffic
&
Transformer-based models are most effective due to attention mechanisms that model long-range dependencies, cross-flow interactions, and domain shifts. Best suited for data-rich scenarios where scalability and flexibility outweigh interpretability concerns.
\\
\hline
\end{tabular}
\end{table}

This matrix highlights that no single generative approach is universally optimal. Increasingly expressive models subsume simpler ones at the cost of higher data requirements, computational overhead, and reduced transparency. Consequently, model selection should be driven by application constraints rather than isolated performance indicators.

\section{Commercial network traffic simulators}\label{sec:commercial}

Traditionally, traffic generators have targeted network protocols and infrastructure, with less emphasis on the applications running above them. This has been especially true with canonical tools like the venerable Iperf~\cite{iperf}, which is often employed to test end-to-end transport protocols, or hardware-based solutions such as IXIA\cite{ixia}, Spirent\cite{spirent}, or XENA\cite{xena} which are designed for use cases such as performing RFC 2544 compliant device tests~\cite{rfc2544}. Traditionally, the division of work between software and hardware-based solutions presented a trade-off, where software tools often fell short in their ability to reach the full core network line rate, which is often achieved by specialized hardware.

However, as networks, applications, operating systems, and COTS (commercial off-the-shelf) hardware have evolved, so too have the associated evaluation tools. In recent years, there has been a noticeable shift towards the development of both software and hardware solutions that enable the creation of flexible traffic generators at a relatively low cost. This shift has been driven in part by advancements in kernel bypassing technologies, such as DPDK (Data Plane Development Kit) and eBPF (Extended Berkeley Packet Filter), as well as the widespread availability of programmable Network Interface Cards (NICs), including FPGA (Field-Programmable Gate Array), IPU (Infrastructure Processing Unit), and smartNICs. 

These technological advancements have significantly impacted the networking and distributed systems communities, leading to the development of more versatile tools. On one hand, these tools are capable of achieving performance levels comparable to those of special-purpose hardware solutions when used with commodity network cards~\cite{moongen-imc2015, netfpga, trex}. On the other hand, they offer tailored traffic generation for specific workloads, such as key-value stores~\cite{DBLP:conf/isca/ZhangMMT16, lancet-atc19, DBLP:conf/eurosys/LeverichK14, microbursts-imc17, cloud-perf-nsdi18}. This dual capability has opened new possibilities for network testing, allowing for more precise and application-specific evaluations, which were previously challenging to perform with older, less flexible tools. 

The landscape of traffic generation has been further enriched by comprehensive surveys such as that by Adele \textit{et al.}~\cite{adeleke2022network}, who explored the functional behaviors of traffic generators across academia and industry. Rather than focusing solely on raw performance metrics, this study categorized traffic generators into seven distinct classes, ranging from "Constant or Maximum Throughput Generators" to "Script-Driven Traffic Generators." Their analysis also spotlighted the top ten most popular or cited tools between 2006 and 2018, including widely recognized names like Iperf~\cite{iperf}, Scapy~\cite{scapy}, MoonGen~\cite{moongen-imc2015}, or Tcpreplay, many of which remain highly relevant today.

A significant trend observed in recent years is the evolution of legacy traffic generators from standalone tools to integral components within broader performance evaluation frameworks. This shift is especially pronounced in commercial network monitoring and evaluation software suites, where traffic generation functions as one of many tools alongside capabilities such as network tomography and periphery scanning. Examples include SolarWinds Network Performance Monitor~\cite{solarwinds}, Dynatrace~\cite{dynatrace}, or NetScanTools Pro~\cite{netscantools}  all of which incorporate basic traffic generation features. These implementations often rely on legacy traffic generators while integrating them with canonical networking utilities like traceroute~\cite{traceroute} or nmap~\cite{nmap}.

Moreover, the flexibility and performance of modern traffic generators have broadened the scope of testing possibilities. No longer confined to infrastructure-centric evaluations, these tools now support complex, high-fidelity simulations of real-world application scenarios. As network environments grow increasingly dynamic with the proliferation of cloud computing, IoT devices, and edge computing, the demand for adaptive and precise evaluation tools continues to rise. 


\begin{table*}[]
\scriptsize
    \centering
    \caption{Summary of commercial traffic generators}
    \label{table:comm_summary}
    \begin{tabular}{|P{2cm}|P{1cm}|P{4.8cm}|P{1.5cm}|P{1.25cm}|P{5cm}|}
    \hline
    \textbf{Work}& \textbf{Type}&\textbf{Protocol constraints}&\textbf{Cost}& \textbf{Usability}& \textbf{Main application}\\
    \hline
    IXIA~\cite{ixia} & Hardware & Datalink layer to transport layer by default, additional modules for above & Expensive & GUI & Compliance and System Engineering\\
    \hline
    Spirent~\cite{spirent}& Hardware & Datalink layer to transport layer by default, additional modules for above & Expensive & GUI & Compliance and System Engineering\\
    \hline
    XENA~\cite{xena} & Hardware & Datalink layer to transport layer by default, additional modules for above& Expensive & GUI & Compliance and System Engineering\\
    \hline
    PROPSIM~\cite{KeysightPROPSIM2023} & Hardware & Physical layer (Real-world RF channel) & Expensive & GUI and script & RF channel emulation for wireless device testing\\
    \hline
    CMW series~\cite{RohdeSchwarzCMW2023} & Hardware & Physical layer (Wireless protocol signaling)  & Expensive & GUI and script & Multi-standard wireless testing and network emulation\\
    \hline
    Trex~\cite{trex}& Software & Datalink layer to transport layer (stateless) & Low/Open source& GUI and script & Performance and management\\
    \hline
    SolarWinds~\cite{solarwinds} & Software& Datalink layer to transport layer & Medium& GUI and script & Performance and management\\
    \hline
    Dynatrace~\cite{dynatrace} & Software& Datalink layer to transport layer  by default, need to program for more& Medium& GUI and script & Performance and Management\\
    \hline
    NetScanTools~\cite{netscantools} & Software& Datalink layer to transport layer & Medium& GUI & Development and penetration testing \\
    \hline
    OSINATO~\cite{ostinato} & Software& Datalink layer to transport layer & Low/Open source& GUI script & Development and penetration testing \\
    \hline
    Scapy~\cite{scapy} & Software & Datalink layer to application layer& Low/Open source& script & Development and penetration testing\\
    \hline
    Matlab toolbox~\cite{MathWorksWirelessToolboxes2024} & Software & Physical layer & Expensive & GUI and script & System design and waveform generation\\
    \hline
    GNU Radio~\cite{Blossom2004} & Software & Physical layer& Low/Open source& GUI and script & Development and testing\\
    \hline
    QuaDRiGa~\cite{Jaeckel2014} & Software & Physical layer& Low & script & 3D radio channel impulse‑response modeling \\
    \hline

\end{tabular}
\end{table*}

In order to reflect this de-facto specialization of tools, we present in Table~\ref{table:comm_summary} an overview of some of the most popular industry-used traffic generators. In particular, we propose to classify those tools under three broad categories: compliance and system engineering, performance evaluation and management, and development and penetration testing. As a result, we can see  that due to various regulations, traffic generators for compliance and system engineering tend to be dominated by hardware-based solutions as they made sure to pass numerous certifications for a given piece of equipment. In particular, we note that nowadays, this category is still dominated by tools such as IXIA\cite{ixia}, Spirent\cite{spirent}, or XENA\cite{xena}. 


In the other category of traffic generator applications, the performance and network management, we note the presence of one legacy fully dedicated traffic generator application, Trex~\cite{trex}, and two traffic generator frameworks, SolarWinds and Dynatrace. Considering Trex, while it started more as a testing tool, its integration with the rest of the CISCO infrastructure, such as the umbrella framework({\url{https://umbrella.cisco.com/}), made this tool use more in performance and management application with attention to performance analysis.

Then, we present in the last identified traffic generator category, namely the development and penetration testing, three tools that are associated with this category. In particular, we mention in Table~\ref{table:comm_summary}, two tools offering a modular testing capability with an easy-to-use Graphical User Interface (GUI) in NetScannTools and Osinato~\cite{ostinato} frameworks. Also, we mention a tool that has been gaining more attraction over the years with Scapy, where the user can easily analyze, capture, and replay any type of traffic as long as performance and high throughput are not mandatory due to its python implementation. 

In addition to traditional packet-level traffic generators, the wireless and physical-layer domain is supported by specialized signal and channel simulators. MATLAB’s wireless toolboxes provide functions for generating standards-compliant RF waveforms and channel models \cite{MathWorksWirelessToolboxes2024}. 
Open-source Software Defined Radio (SDR) platforms like GNU Radio~\cite{Blossom2004} likewise enable custom wireless signal generation and reception. GNU Radio provides modular signal processing blocks for real-time implementation of Wi-Fi, LTE, or other wireless protocols, and is widely used in both academic and commercial wireless research. Another example is QuaDRiGa~\cite{Jaeckel2014} (Quasi-Deterministic Radio channel GenerAtor), a MATLAB toolbox for creating realistic 3GPP-compliant MIMO channel impulse responses for system-level simulations. These tools output physical-layer waveforms and measurements (e.g. per-subcarrier channel estimates, RSSI/Reference Signal Received Power[RSRP]) rather than just packet traces, complementing higher-layer traffic generators.

On the hardware side, commercial channel emulators and wireless testers provide realistic over-the-air signal generation. For instance, Keysight’s PROPSIM~\cite{KeysightPROPSIM2023} channel emulation systems can reproduce multipath fading, Doppler shifts, pathloss, and noise in a lab environment to test 5G, Wi‑Fi, and other wireless devices. Similarly, Rohde and Schwarz~\cite{RohdeSchwarzCMW2023} offers wireless network emulators (e.g. the Rohde \& Schwarz [R\&S] CMW500/CMX500 series) that generate multi-band LTE/5G/Wi‑Fi signaling with configurable MIMO channel models. These platforms produce actual RF waveforms and report physical layer metrics (RSSI/RSRP, CSI, Error Vector Magnitude [EVM], etc.) under controlled channel conditions. Together, these physical-layer tools extend the scope of traffic generation by embedding real-world wireless channel effects into network testing.
    

\section{Challenges and Future Directions}
\label{sec:challenges_and_future_directions}
\subsection{Privacy and Utility Trade-off}

\textbf{Challenge:} Tradeoff between privacy and utility is a main challenge in network traffic generation. Real network data often contains sensitive information such as IP addresses, and communication patterns. making privacy protection essential. However, applying mechanisms such as DP, often degrade the quality of the synthetic data, leading to reduced performance in downstream tasks. This opens up the privacy-utility trade-off. DP introduces noise into the data to mask individual entries, but this can distort key traffic patterns. The challenge is to find the right balance between the level of noise (controlled by the privacy parameter \(\epsilon\)) and the ability of the retaining useful properties in synthetic data. Fine-tuning \(\epsilon\) requires a deep understanding of the privacy risks and application requirements. 
Some preliminary work has been done in \cite{yin2022practical, lin2020using} to address the privacy-utility trade-off by incorporating DP into the GAN generation, nevertheless requires further improvements. 

\textbf{Future Direction:} Future privacy-preserving traffic generators can leverage advanced generative models such as DMs and Transformers, incorporating noise at more context-aware levels (e.g., feature-specific perturbations). Multimodal generative models may allow selective obfuscation of sensitive features while preserving utility in others. Beyond DTs, one may also envision the possibility of networks of synthetic traffic sources ( a DT without a physical counterpart) generating and inserting traffic to physical networks for tasks such as camouflaging or protecting privacy.  With the rise of Large Language Models (LLMs) based on GPT architectures across various domains, users increasingly rely on cloud-based models, necessitating the filtering of private data before transmission to ensure privacy and security.

\subsection{Adhering to Protocol Standards and Realistic Conditions}

\textbf{Challenge:}  Synthetic network data has to be representative of the  timestamp and volume distributions of  underlying networks, protocols and conditions they claim to emulate,  and  must not be in violation of conditions imposed by the strict protocol standards that govern data integrity and transmission quality. Ensuring that synthetic traffic adheres to the protocols is critical for realistic and useful synthetic data. Over 80\% of the work does not consider developing network traffic generation that adheres to protocol standards. This has been addressed to a certain level in \cite{jiang2024netdiffusion, jiang2023generative} by integrating ControlNet into DM for data generation to adhere to protocol standards and performing a post-processing step to correct the images to ensure that they adhere to the protocol. However, there still exist certain challenges to be addressed.

\textbf{Protocol Compliance:} Network protocols define limits on packet sizes, acceptable payload structures, header formats, and timing between packets. For instance, protocols like TCP/IP and UDP have maximum packet size constraints and specific sequencing requirements. Synthetic data must comply with these standards to ensure usability for applications like simulation and stress testing of network systems.

\textbf{Timestamp Synchronization and Temporal Coherence:} Network data is highly time-sensitive, with applications that rely on precise timestamps and sequence orders for accurate simulations. Synthesis algorithms must handle timestamps and ensure that events in the synthetic dataset mirror the temporal coherence found in real network traffic. This challenge becomes even more pronounced in applications like IDS, where real-time analysis of packet sequences is essential.

Further, it is nearly impossible to collect data for different realistic conditions to statisfy every possible scenarios. Network traffic synthesis will be a major player to alleviate this problem. This is further discussed in section \ref{subsec:unseen_scenarios}.

\textbf{Future Direction:} Future models could explicitly incorporate protocol rules as conditional constraints or even embed Request for Comments (RFC)-level formal specifications. Additionally, timestamp-aware generative models trained using temporal coherence losses can better preserve the temporal order of packets. These improvements are crucial for generating data that conforms to real-world constraints beyond mere protocol adherence. Moreover, realistic simulation environments such as 5G DTs can serve as testbeds for validating protocol- and time-aware data generation under operational conditions. Embedding traffic generators in AI-based network design frameworks like Ericsson’s 5G DT~\cite{NVIDIA2022} opens new avenues for training, evaluation, and deployment of realistic synthetic traffic under evolving network dynamics.

\subsection{Capturing Complex Network Behaviors}

\textbf{Challenge:} Network traffic exhibits highly complex behaviors, shaped by dynamic routing, load balancing, and bursty traffic flows. Capturing these nuances in synthetic data remains a significant challenge. Some efforts have addressed this to an extent.For example, \cite{kattadige2021videotrain, madarasingha2022videotrain++, sivaroopan2023synig} use packet binning to compress time series traces while retaining key features, and \cite{sivaroopan2023synig, sivaroopan2024netdiffus} time-series to 2D imaging to capture complex temporal correlations in packet-level data. However, despite these advancements, further challenges remain in accurately modeling the intricate characteristics of network traffic.

\textbf{Traffic Diversity:} Network traffic varies greatly depending on the type of applications, devices, and protocols in use. Synthetic data generators need to be flexible enough to capture this diversity, from low-latency traffic such as video streaming to high-throughput traffic such as large data transfers. However, existing synthesis methods often fail to replicate these variations effectively, leading to over-simplified or unrealistic synthetic traffic patterns.

\textbf{Temporal Dynamics and Seasonality:} Network traffic is not static; it changes over time due to user behavior and network conditions. Ensuring that synthetic data models these temporal dynamics, including short-term fluctuations and long-term seasonality, is a major challenge.

\textbf{High-dimensional Feature Spaces:} Network traffic data is often high-dimensional, encompassing a wide range of features such as packet sizes, inter-arrival times, protocol flags, and source-destination IP pairs. Capturing correlations and dependencies across these features in a synthetic dataset is difficult. Synthetic data may either over-simplify these relationships, lose important interactions, or become too complex to be useful for downstream tasks.

\textbf{Future Direction:} Future commercial traffic generators could benefit from advanced generative techniques like DMs and Transformer-based architectures (GPTs), which, pre-trained on large datasets, offer powerful sequence generation capabilities. 
By bridging the gap between the LLM’s word embedding space and the embedding space of time-series tokens or packet flows, GPTs can be fine-tuned to generate realistic and contextually rich network traffic patterns. Similarly, DMs, known for their iterative refinement of data from noise, offer an exceptional ability to model intricate data distributions and generate synthetic data with high fidelity. These models could be effective in mimicking the statistical and temporal properties of network traffic, capturing both short-term variations and long-term dependencies.

Another way to capture complex patterns is through multimodal deep generative models that multiple data types (e.g., combine text, images, network metadata (e.g., headers, flow statistics), and raw RF signals). DMs such as Stable Diffusion can be conditioned on network topology diagrams or textual descriptions of attack scenarios to generate structured protocol-compliant synthetic traffic. Similarly, transformer (e.g. Contrastive Language-Image Pretraining (CLIP)~\cite{radford2021learning}) models can align textual representation of network behavior with images or other modalities, allowing for more context-aware traffic synthesis that is more interpretable and adaptive. GPTs can extend their sequence modeling capabilities beyond packet flows by incorporating visual and textual network telemetry, enabling cross-modal learning for real-time network traffic simulation. The synergy of these techniques in commercial applications could enable scalable and precise traffic generation solutions, addressing long-standing challenges such as data imbalance, privacy preservation, and real-time traffic simulation. By leveraging these SOTA multimodal generative models, the next generation of commercial traffic generators could offer enhanced capabilities, transforming how synthetic network traffic is created and utilized for cybersecurity, network optimization, and anomaly detection.

\subsection{Evaluating Synthetic Data Quality}

\textbf{Challenge:} Another key challenge lies in developing standardized metrics for evaluating the quality of synthetic network traffic data. While there are several metrics such as statistical similarity and performance on downstream tasks, there is no one-size-fits-all solution. There are many papers ~\cite{kattadige2021360norvic, kattadige2021seta++, sivaroopan2023synig, sivaroopan2024netdiffus} attempting to provide high-quality synthetic data, however, there are challenges yet to be solved.

\textbf{Lack of Benchmark Datasets:} Unlike other fields such as computer vision or NLP, there are relatively few publicly available, standardized benchmark datasets for synthetic network traffic evaluation. This makes it difficult to compare the performance of different synthesis approaches, especially for privacy-preserving techniques.

\textbf{Generalization vs. Overfitting:} Striking the right balance between generalization and overfitting in synthetic data generation remains a major issue. Overfitting to the real traffic can lead to privacy risks, while excessive generalization may result in synthetic data that fails to capture the essential characteristics of the network traffic, reducing its utility.

\textbf{Task-Specific Evaluation:} The effectiveness of synthetic data can vary significantly depending on the specific task it is used for, such as traffic classification, anomaly detection, or intrusion prevention. This calls for task-specific evaluation metrics, but defining these metrics can be complex and may require different evaluation frameworks for each application.

\textbf{Future Direction:} A unified evaluation toolkit combining statistical, task-specific, and privacy-preserving metrics is needed.
Synthetic traffic benchmarks, similar to ImageNet~\cite{deng2009imagenet} in the computer vision domain, would facilitate standardized and consistent comparisons across traffic generation methods. Also, interpretable metrics linking traffic properties with task-level performance can help building better generation models.


\subsection{Computational Complexity and Scalability}

\textbf{Challenge:} As networks expand in size and complexity, the computational demands of generating and evaluating synthetic data grow accordingly. This also poses a particular challenge for real-time network traffic generation on edge devices, where computational resources are limited. To address this, the authors of \cite{li2024lightweight} introduced lightweight DMs designed for efficient data generation on edge devices. However, further challenges remain in balancing computational efficiency with the accuracy and fidelity of the generated traffic.

\textbf{Scalability of Data Generation:} Synthesizing data mimicking large scale networks (with sheer volume of data) requires significant computational resource.
Many existing synthesis methods struggle to scale effectively, leading to bottlenecks in generating synthetic data for large network environments.

\textbf{Real-time Synthesis:} In some applications, synthetic data may need to be generated in real-time to support testing and analysis under current network conditions. Achieving real-time synthesis is particularly challenging when attempting to replicate complex traffic patterns while maintaining high fidelity and privacy guarantees.

\textbf{Future Direction:} Future directions include the development of highly parallelized generation pipelines, federated learning and model distillation techniques for real-time inference. 
Edge-optimized models can be tailored to resource constraints while preserving core behavioral patterns. Integration into systems like Ericsson’s 5G DT~\cite{NVIDIA2022}, simulated in NVIDIA Omniverse, opens possibilities for real-time simulation and testing under realistic and evolving network conditions, supporting next-generation AI-driven network operations. Further, LLMs in many applications can also be replaced with evolving Small Language Models (SLMs).

\subsection{Synthetic Data Generation for Unseen Network Scenarios}
\label{subsec:unseen_scenarios}

\textbf{Challenge:} Synthesizing network data that represents unseen scenarios is crucial for evaluating system resilience and preparedness in various environments. Generative models can simulate unseen network conditions by learning from available datasets and extrapolating patterns for unobserved scenarios, such as extreme network loads, and rare cyberattacks etc. Additionally, when real-world data for specific platforms or protocols is unavailable, synthetic data generation can incorporate domain knowledge to approximate these environments. However, the key challenge lies in ensuring that the generated data remains both valid and useful for downstream tasks.

\begin{figure}[h!]
\centering
\includegraphics[width=0.8\linewidth]{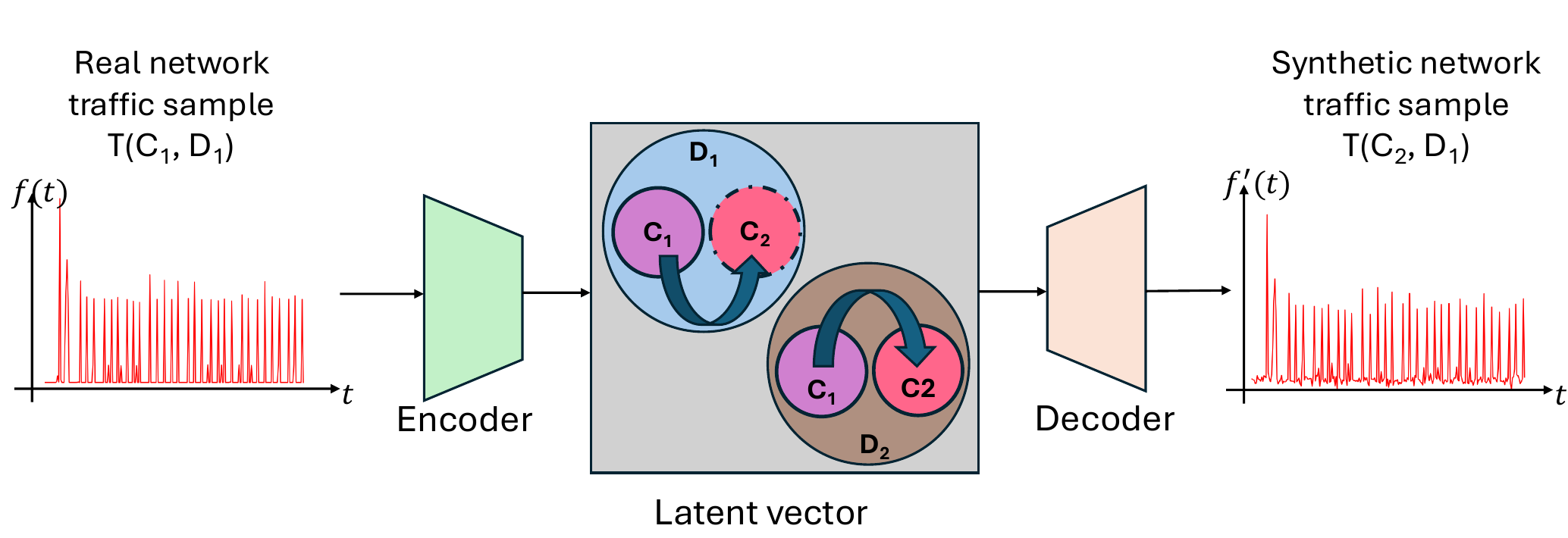}
\caption{Scenario-based data generation: VAE}
\label{fig:unseen_vae}
\end{figure}

\textbf{Future Direction:} One strategy is to follow \textit{Neural Style Transfer (NST)}~\cite{gatys2015neural}, which blends the content of one image with the style of another. NST is based on the intuition that the \textit{content} (structure and high-level features) and \textit{style} (textures, patterns, and colors) of an image can be extracted using different layers of a deep convolutional network and that optimizing a new image to match the content features of one image and the style features of another generates a result that retains the structure of the content image while adopting the artistic characteristics of the style image. 
To follow a similar approach, we first define two attributes for each network trace, \textit{Content} which refers to the actual content/payload carried by the traffic trace (such as a video stream, or website), and \textit{Domain} which refers to the conditions of the network which affects the traffic. For instance, when considering a network trace for a user streaming videos, the specific video title would be the content and the domain could be the specific network the user is connected to (wired or wireless) or the specific geographical location. The notation $T_n (C_x, D_y)$ represents a traffic trace corresponding to content $x$ and domain $y$. 
For example, consider a scenario where traffic traces with content $C_1$ from domain $D_1$, and traffic traces with content $C_2$ from domain $D_2$ are available, but traffic traces with content $C_2$ from domain $D_1$ is not available and needs to be generated. Analogous to NST, if we can extract content-specific features from $T_2 (C_2, D_2)$ and domain-specific features from $T_1 (C_1, D_1)$, these features can be combined to generate a new sample $T_3 (C_2, D_1)$ representing $ C_2$ and $ D_1$. 

While the layers of deep convolutional networks for extracting content and style in images are well established, no comparable approach has, to our knowledge, been studied for network traffic traces, leaving it an open research problem. However, P. Smolovic \textit{et al}.~\cite{tripletvinet} introduced the use of triplet loss to extract video-specific (content) features from network traffic traces, independent of the platform on which the video was streamed (domain). Based on this, we propose that using a methodology leveraging contrastive loss functions (e.g., triplet loss), to extract both content and domain-specific features from network traffic traces is feasible. Once the content and domain-specific features are extracted, the same approach as NST can be followed to generate the new sample by optimizing a random input to match the extracted content and domain. 

An alternative is to use an encoder–decoder architecture such as a VAE, which structures the latent space into clusters by domain and content type. As shown in Figure~\ref{fig:unseen_vae}, the model can form distinct domain clusters, with further subdivisions by content. Once trained, the latent vector for a traffic trace could be extrapolated in the latent space to generate a trace with a different content or domain. For example, Figure~\ref{fig:unseen_vae} shows a scenario where the model is trained on traces with content $C_1$ from both domains and $C_2$ from $D_2$ only. Accordingly, it will learn both $C_1$ and $C_2$ clusters in $D_2$ and only $C_1$ cluster in $D_1$. Once trained a trace $T(C_1, D_1)$ can be extrapolated in the latent space to create $C_2$ in $D_1$ to match context $C_2$ based on the relationship between $C_1$ and $C_2$ in $D_2$. Then the decoder can generate a realistic network traffic sample $T(C_2, D_1)$ that represents context $C_2$ and domain $D_1$.

\section{Conclusion}
\label{sec:conclusion}

We presented a comprehensive review of synthetic network traffic generation using different mechanisms, mainly focusing on deep learning–based approaches. We discussed widely used statistical methods and commercial simulation tools for network traffic generation, as well as the different traffic types employed in the generation process. Furthermore, we provided a comparative analysis of these mechanisms through a novel approach based on LLMs. This structured classification offers insights into the strengths, limitations, and practical applications of each approach, showing the evolution of the field and how these methods are applied to synthesize diverse scenarios. Finally, we highlighted the existing challenges in network traffic generation and with future directions to guide the further research. 

Several open challenges remain for the community, including balancing privacy with utility, adhering to protocol standards and realistic conditions, capturing diverse and complex behaviors, and developing standardized benchmarks. Scalability and efficiency for real-time synthesis, as well as generating realistic traffic for rare scenarios, are also pressing needs. Addressing these issues will drive the next wave of innovation, enabling more accurate, scalable, and trustworthy models for applications in cybersecurity, network optimization, and system testing.

\section{Acknowledgement}
\label{sec:ack}

This research was partially funded by the Australian Research Council Industrial Transformation Research Hub for Future Digital Manufacturing (IH230100013).

\balance
\bibliographystyle{IEEEtranS}
\bibliography{99_bib}

\end{document}